\documentclass[%
 aip,
 superscriptaddress,
 amsmath,amssymb,
 reprint,%
]{revtex4-2}

\usepackage{graphicx}
\usepackage{dcolumn}
\usepackage{bm}
\usepackage{xcolor}
\begin{document}

\preprint{AIP/123-QED}

\title{Dynamical States and Bifurcations in Coupled Thermoacoustic Oscillators}

\author{Sneha Srikanth}
 \affiliation{%
 Department of Mechanical Engineering, Indian Institute of Technology Madras, Chennai 600036, India
 }%
\author{Samadhan A. Pawar}%
\affiliation{%
 Department of Aerospace Engineering, Indian Institute of Technology Madras, Chennai 600036, India
}%
 \email{samadhanpawar@ymail.com}
\author{Krishna Manoj}%
\affiliation{%
Department of Mechanical Engineering, Massachusetts Institute of Technology, Cambridge, MA 02139, USA
}%
\author{R. I. Sujith}%
\affiliation{%
 Department of Aerospace Engineering, Indian Institute of Technology Madras, Chennai 600036, India
}%

\date{\today}

\begin{abstract}
The emergence of rich dynamical phenomena in coupled self-sustained oscillators, primarily synchronization and amplitude death, has attracted considerable interest in several fields of science and engineering. Here, we present a comprehensive theoretical study on the manifestation of these exquisite phenomena in a reduced-order model of two coupled Rijke tube oscillators, which are prototypical thermoacoustic oscillators. We characterize the dynamical behaviors of two such identical and non-identical oscillators by varying both system parameters (such as the uncoupled amplitudes and the natural frequencies of the oscillators) and coupling parameters (such as coupling strength and coupling delay). The present model captures all the dynamical phenomena -- namely synchronization, phase-flip bifurcation, amplitude death, and partial amplitude death -- observed previously in experiments on coupled Rijke tubes. By performing numerical simulations and deriving approximate analytical solutions, we systematically decipher the conditions and the bifurcations underlying the aforementioned phenomena. The insights provided by this study can be used to understand the interactions between multiple cans in gas turbines combustors and develop suitable control strategies to avert undesirable thermoacoustic oscillations in them.
\end{abstract}

\maketitle

\begin{quotation}
The interactions between two coupled self-sustained oscillators can lead to a myriad of dynamical phenomena. For example, depending on the scenario, the coupled oscillators may adjust their rhythms and consequently synchronize, or both the oscillators may cease to oscillate, resulting in amplitude death. At times, the coupled oscillators may attain partial amplitude death, i.e., the oscillations in only one of the oscillators may be suppressed, while the other exhibits comparatively large amplitude oscillations. Though the aforementioned phenomena have been demonstrated recently in experiments on two thermoacoustic oscillators coupled acoustically using a single connecting tube, they are yet to be corroborated through modeling. Moreover, their underlying bifurcations are yet to be investigated. In the present study, we consider a model of two thermoacoustic oscillators subjected to delay coupling.
Through numerical and analytical techniques, we throw light on how parameters such as coupling strength, coupling delay, heater power, and frequency mismatch affect the bifurcations leading to synchronization, amplitude death, and partial amplitude death in the system. 
\end{quotation}

\section{\label{sec:level1}Introduction \protect}
Coupled nonlinear oscillators have garnered considerable interest due to their pervasive applications in domains extending from biological to engineering systems \cite{winfree1967biological,van1985observation,roy1994experimental, manrubia2004emergence, jenkins2013self, zou2021quenching}. Populations of coupled oscillators can exhibit a wide variety of exquisite phenomena depending on the nature of coupling between them \cite{atay2004delays, balanov2008synchronization, bera2017chimera, boccaletti2018synchronization, manoj2021experimental}. The most widely studied phenomenon among them is synchronization, which refers to the adjustment of rhythms of coupled oscillators due to the mutual interactions between them \cite{strogatz2004sync, pikovsky2003synchronization}. These interactions can sometimes lead to complete suppression of all oscillations in the system; i.e., all the constituent oscillators reach a homogeneous steady state. This phenomenon, which was first discovered by Rayleigh \cite{rayleigh1896theory}, is referred to as amplitude death (AD) \cite{mirollo1990amplitude}. The occurrence of AD has been demonstrated experimentally and theoretically in many systems with different coupling schemes including delay, dissipative, and conjugate couplings \cite{saxena2012amplitude, koseska2013oscillation, lakshmanan2011dynamics, zou2021quenching}. A system of coupled oscillators can also attain partial amplitude death (PAD), a dynamical state wherein some of the oscillators in the system are completely damped (or exhibit very small amplitude periodic oscillations) while the others exhibit comparatively large amplitude limit cycle oscillations \cite{atay2003total}. This state is generally observed in systems of coupled non-identical oscillators \cite{koseska2013oscillation}. 

Traditionally, the routes to various dynamical states in coupled oscillators have been studied only by varying their coupling parameters (such as coupling strength and coupling delay). For example, several studies have varied coupling strength to investigate the route to synchronization and amplitude death in coupled Kuramoto oscillators \cite{acebron2005kuramoto} and Stuart-Landau oscillators \cite{saxena2012amplitude,zou2021quenching}, respectively. Studies have also characterized the effect of coupling delay on the occurrence of phase-flip bifurcation (PFB), which is the abrupt transition of a coupled system from a state of in-phase synchronization (IP) to a state of anti-phase synchronization (AP) or vice-versa \cite{prasad2006phase,karnatak2010nature,manoj2018experimental}. However, recent studies by Dange \textit{et al.} \cite{dange2019oscillation} and Premraj \textit{et al.} \cite{premraj2021effect} indicate that, in addition to coupling parameters, variation in system parameters, such as the amplitude and the natural frequency of oscillators, significantly affects the dynamics of coupled systems. Although previous studies have considered the effect of varying the natural frequencies of oscillators on the dynamics of coupled systems \cite{aronson1990amplitude, reddy1999time}, the effect of change in the amplitude of the oscillators on their coupled behavior has not been investigated extensively. 

The occurrence of large amplitude self-sustained oscillations can have disastrous consequences to systems in real life. These consequences include structural damage to combustors due to thermoacoustic instabilities \cite{lieuwen2005combustion, culick2006unsteady}, destruction of aircraft wings due to fluttering \cite{garrick1981historical}, wobbling and collapse of bridges \cite{strogatz2005crowd,green2006failure}, spread of epidemics \cite{duncan1997dynamics,rypdal2021tipping}, crashes in financial markets \cite{frankel2008adaptive}, and so on. These oscillatory instabilities possess widely different amplitudes and natural frequencies. Hence, in order to effectively control them, it is vital to understand how changes in the inherent system parameters of the oscillators can alter their coupled behavior. More specifically, we need to decipher the nature of bifurcations underlying the transitions between different dynamical states on the variation of system parameters in coupled oscillators. Towards this end, we investigate how system parameters and coupling parameters affect the dynamical behavior of a model of coupled horizontal Rijke tube oscillators \cite{balasubramanian2008thermoacoustic}, using bifurcation analysis and synchronization theory. 

The Rijke tube is a classical example of a thermoacoustic oscillator, consisting of a simple tube open at both ends with a heat source present inside \cite{rijke1859lxxi}. A thermoacoustic oscillator refers to a confined system wherein the positive feedback between the heat release rate fluctuations of the heat source and the acoustic field of the system gives rise to large amplitude self-sustained tonal sound waves. The occurrence of these high amplitude acoustic oscillations is known as thermoacoustic instability \cite{lieuwen2005combustion, sujith2020complex,sujith2021book}. The presence of thermoacoustic instability has detrimental effects on the structural integrity of gas turbine combustors and rocket engines \cite{lieuwen2005combustion,poinsot2017prediction,culick2006unsteady}. Many mitigation strategies have been developed over the years to control thermoacoustic instability in individual combustion systems \cite{mcmanus1993review,lieuwen2005combustion,huang2009dynamics,sujith2020complex}. However, most practical gas turbines, such as can type or can-annular type combustors, consist of multiple combustion systems which interact with each other and lead to the simultaneous occurrence of thermoacoustic instabilities in more than one system \cite{kaufmann20083d,farisco2017thermo,ghirardo2019thermoacoustics,jegal2019mutual,moon2020mutual,pedergnana2021coupling,moon2021experimental}. It is thus important to understand the complex dynamics resulting from interactions between multiple such coupled thermoacoustic systems and also to develop control strategies to simultaneously mitigate thermoacoustic instabilities in them.

Recently there has been an increased interest to study the effect of mutual coupling on the dynamics of two coupled thermoacoustic systems, theoretically \cite{thomas2018effect,thomas2018noise,hyodo2020suppression,sahay2021dynamics,guan2021low} as well as experimentally \cite{biwa2015amplitude,dange2019oscillation,hyodo2020suppression,sahay2021dynamics}. Biwa \textit{et al.} \cite{biwa2015amplitude} experimentally investigated the occurrence of amplitude death in two thermoacoustic engines that are coupled via both delay and dissipative couplings. Thomas \textit{et al.} \cite{thomas2018effect,thomas2018noise} systematically examined the occurrence of amplitude death in a model of two horizontal Rijke tubes when time-delay and dissipative couplings are added individually and simultaneously. Dange \textit{et al.} \cite{dange2019oscillation} experimentally revealed the existence of amplitude death, phase-flip bifurcation, and partial amplitude death through systematic variation of system and coupling parameters in two horizontal Rijke tubes coupled via a connecting tube. Hyodo \textit{et al.} \cite{hyodo2020suppression} experimentally studied oscillation quenching through double tube coupling in two flame-driven Rijke tube oscillators. Further, Sahay \textit{et al.} \cite{sahay2021dynamics} demonstrated the expansion of AD region in the control parameter space by implementing asymmetric forcing to coupled horizontal Rijke tube oscillators. Although the above studies demonstrate a wide variety of dynamical phenomena in thermoacoustic systems, the mechanisms (or bifurcations) by which such phenomena occur on variation of system parameters or coupling parameters is still not clearly understood. Moreover, the presence of phase-flip bifurcation and partial amplitude death in thermoacoustic systems is yet to be modeled.

Here, we aim to throw light on the route to the myriad of dynamical states observed in previous experiments on coupled horizontal Rijke tube oscillators \cite{dange2019oscillation}. We numerically and analytically show that an interplay between the system parameters and the coupling parameters determines the occurrence of amplitude death in thermoacoustic oscillators. We demonstrate that identical delay coupled Rijke tube oscillators transition between the states of in-phase synchronization and anti-phase synchronization through two routes: i) by undergoing phase-flip bifurcation, or ii) via an intermediate state of amplitude death. When the delay coupled Rijke tube oscillators are non-identical, we also observe a third route via an intermediate state of desynchronization. Additionally, we uncover the presence of partial amplitude death (PAD) in non-identical delay coupled Rijke tube oscillators. On varying the system and coupling parameters, we observe that desynchronized limit cycle oscillations in delay-coupled non-identical Rijke tube oscillators suddenly synchronize and attain PAD en route to amplitude death. 

The rest of the paper is organized as follows. Section \ref{sec:II} provides details of the model for two coupled identical Rijke tube oscillators. In Sec. \ref{sec:IIIA}, we first study the effect of delay coupling on the bifurcations occurring in two Rijke tube oscillators. Subsequently, we numerically and analytically investigate synchronization and amplitude suppression in two delay coupled identical Rijke tube oscillators and explain the presence of AD and PFB in the system. In Sec.~\ref{sec:IIIB}, we demonstrate PAD and desynchronization, and examine the route to amplitude death in delay coupled non-identical Rijke tube oscillators. We finally present our conclusions in Sec.~\ref{sec:IV}.

\begin{figure}[t]
\includegraphics[width=9.0 cm]{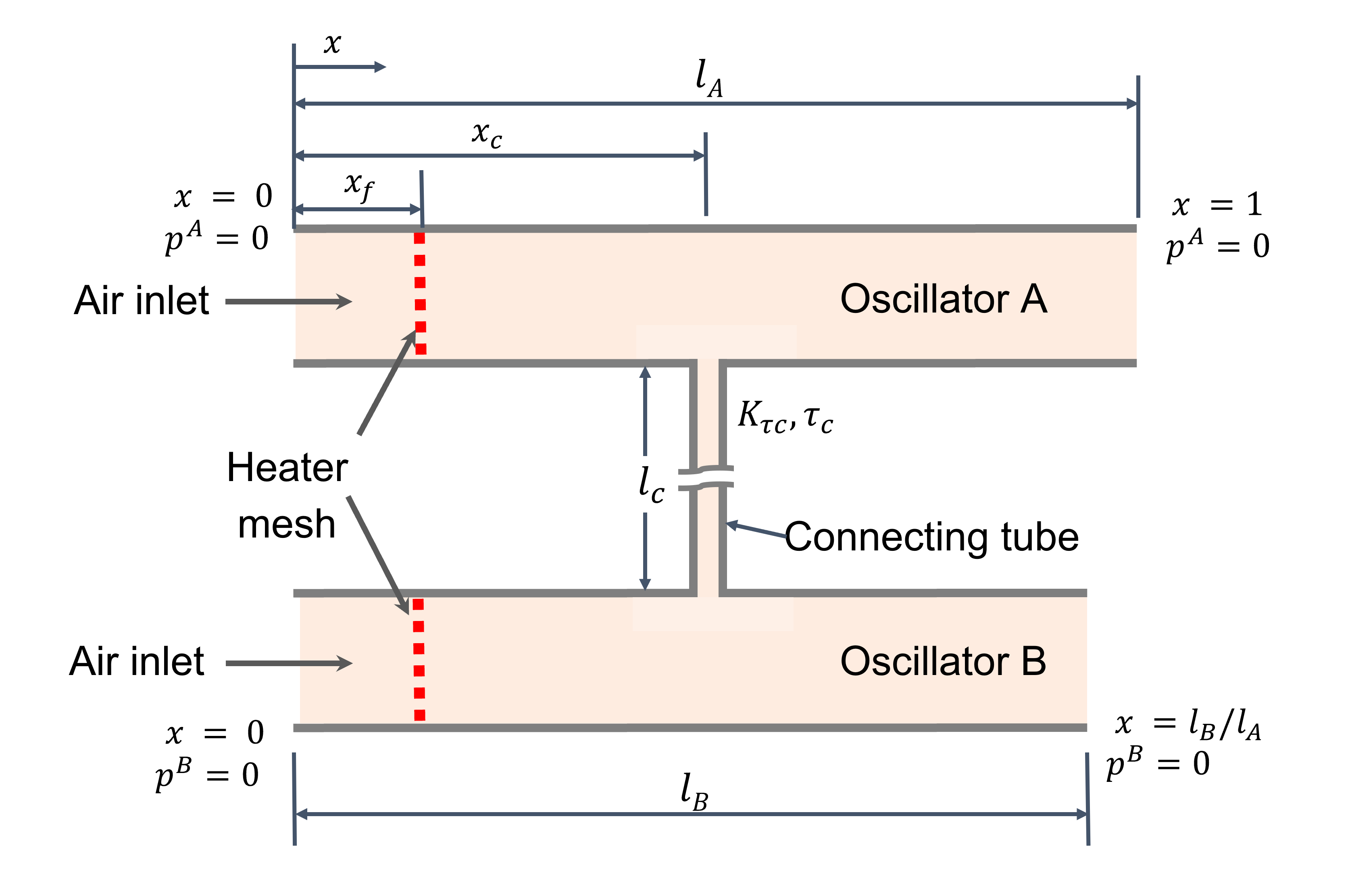}
\caption{\label{fig:1}Schematic diagram of two horizontal Rijke tube oscillators that are delay coupled to each other with coupling strength $K_{\tau c}$ and coupling delay $\tau_c$. Delay coupling can be established in practice by connecting two or more thermoacoustic systems using a connecting tube of length $l_c$.}
\end{figure}

\section{\label{sec:II}Model of coupled Rijke tube oscillators \protect}

In the present study, we consider the horizontal Rijke tube \cite{matveev2003thermoacoustic, gopalakrishnan2014influence} as a nonlinear oscillator. In this system, an electrically heated wire mesh acts as a concentrated heat source. We use the model of the horizontal Rijke tube developed by Balasubramanian and Sujith \cite{balasubramanian2008thermoacoustic}. To model the coupled Rijke tubes [depicted in Fig.~\ref{fig:1}], we first neglect the effects of mean flow (zero Mach number approximation \cite{nicoud2009zero}) and mean temperature gradient in the duct. The resultant non-dimensionalized linearized momentum and energy equations for the acoustic field of the Rijke tubes are as follows:

\begin{align}
    \gamma M \dfrac{\partial u^{A,B}}{\partial t} &+ \dfrac{\partial p^{A,B}}{\partial x} = 0,\label{eq1}\\
    \dfrac{\partial p^{A,B}}{\partial t} +  \gamma M\dfrac{\partial u^{A,B}}{\partial x} &+ \zeta p^{A,B}& \nonumber \\
    &=  (\gamma-1)\dot{Q}^{A,B}(x,t)\delta(x-x_f) 
    \nonumber \\
    &\quad+ C^{A,B}(x,t)\delta(x-x_c).
    \label{eq2}
\end{align}

Here, $u(x,t)$ and $p(x,t)$ are the acoustic velocity and acoustic pressure fluctuations non-dimensionalized by their steady state values $u_0$ and $p_0$, respectively. The superscripts “$A$” and “$B$” denote that the quantities are for oscillators A and B, respectively. $x$ is the distance along the Rijke tubes non-dimensionalized by the length ($l_A$) of oscillator A. The time $t$ is non-dimensionalized by $l_A/c_0$, where $c_0$ is the speed of sound at ambient conditions. $\gamma$ and $M$ are the ratio of specific heats and the Mach number ($M=u_0/c_0$), respectively. $\zeta$ is the damping coefficient. $\dot{Q}(x,t)$ denotes the non-dimensional heat release rate fluctuations per unit area from the heat source, while $C(x,t)$ represents the acoustic coupling between the Rijke tubes. A Dirac delta function is multiplied to $\dot{Q}$ to indicate that the heat source is concentrated at $x_f$, which is the non-dimensional heater location in the Rijke tubes \cite{balasubramanian2008thermoacoustic}. Similarly, $C(x,t)$ is multiplied by a Dirac delta function to indicate that the coupling is located at a position $x_c$ on the Rijke tubes.

We describe the heat release rate fluctuations ($\dot{Q}^{A,B}$) in the individual Rijke tubes using the Heckl's correlation \cite{heckl1990non}:
\begin{align}
    \dot{Q}^{A,B}(x,t) = \frac{2 L_w (T_w - T_0)}{c_0 p_0 S \sqrt{3}}  \sqrt{\pi \lambda_T C_v u_0 \rho_0 r_w}\nonumber\\
    \times \left[ \sqrt{\left\lvert\frac{1}{3} + u^{A,B}(x,t-\tau_h) \right\rvert} - \sqrt{\frac{1}{3}} \right], \label{eq:heckl}
\end{align}
where $r_w$ and $L_w$ are the radius and the length of the heated wire, respectively. $T_w$ and $T_0$ are the temperature of the heated wire and the medium in steady state, respectively. $\rho_0$ is the density of the medium in steady state. $S$ is the cross-sectional area of the duct. $\lambda_T$ is the thermal conductivity, while $C_v$ is the specific heat at constant volume of the medium within the duct. $u^{A,B}(x,t-\tau_h)$ is the acoustic velocity of the oscillators at time $t-\tau_h$. Here, we include the time lag, $\tau_h$, due to the thermal inertia of heat transfer in the medium \cite{lighthill1954response}.

Previous experiments \cite{dange2019oscillation, hyodo2020suppression, jegal2019mutual, moon2019combustion, biwa2015amplitude} suggest that acoustic waves take a finite time to propagate between two systems through the connecting tube. To capture this delayed interaction between the Rijke tubes, we use time-delay coupling \cite{thomas2018effect, sahay2021dynamics, guan2021low}, described by the following expression:
\begin{align}
    C^{A,B}(x,t) = K_{\tau c} \left[p^{B,A}(x, t - \tau_c) - p^{A,B}(x,t) \right], \label{eq:coupling}
\end{align}
where $K_{\tau c}$ is the coupling strength and $\tau_c$ is the coupling delay. This coupling indicates that the interactions between the Rijke tubes is determined by the pressure difference between them. Though there are various other forms of coupling \cite{zou2021quenching}, we choose the above coupling since it qualitatively captures most of the experimental results, as we demonstrate in Sec.~II of the Supplementary Material. Furthermore, this simple form also gives us an understanding of the essential features of the coupling that affect the dynamical behavior of the coupled Rijke tubes. 

To simplify the partial differential equations, $u^{A,B}$ and $p^{A,B}$ are expressed in terms of their Galerkin modes as follows \cite{subramanian2010bifurcation,sahay2021dynamics}:

\begin{align}
     u^A(x,t) &=\sum_{j=1}^{N}{U^A_j(t)\cos(k_jx)},
     \label{eq:1A}
\\
p^A(x,t) &=\gamma M\sum_{j=1}^{N}{P^A_j(t)\sin(k_jx)},
\label{eq:2A}
\\
u^B(x,t) &=\sum_{j=1}^{N}{U_j^B(t)\cos(k_jx/r)},
     \label{eq:1B}
\\
p^B(x,t) &=\gamma M\sum_{j=1}^{N}{P_j^B(t)\sin(k_jx/r)},
\label{eq:2B}
\end{align} 
where $k_j=j\pi$ refers to the non-dimensional wave number of the $j$\textsuperscript{th} mode. $U_j^{A,B}(t)$ and $P_j^{A,B}(t)$ capture the temporal variation of the $j$\textsuperscript{th} modes of $u^{A,B}$ and $p^{A,B}$, respectively. Here, $r=l_B/l_A$ is the ratio between the length of the Rijke tubes. Since the Rijke tubes are open at both ends, the total pressures ($p_t$) at the boundaries is equal to the ambient pressure ($p_0$). Thus, the acoustic pressure fluctuations, $p = p_t - p_{0}$, at the boundaries are zero. Hence, the Galerkin modes are chosen such that the boundary conditions $p^A(0,t)=p^A(1,t)=0$ in oscillator A and $p^B(0,t)=p^B(r,t)=0$ in oscillator B are satisfied. 

Employing the Galerkin technique after substituting Eqs.~\eqref{eq:heckl}-\eqref{eq:1A} in Eqs.~\eqref{eq1} and \eqref{eq2}, we obtain the governing equations for the delay coupled Rijke tubes A and B as:

\begin{gather}
    {\dot{U}}^A_j+k_j P^A_j =0,
    \label{eq:gov1}
\end{gather}
\begin{align}
{\dot{P}}^A_j &+ 2\zeta_j\omega_j P_j^A - k_jU_j^A
\nonumber\\
&= W^A\Bigg(\sqrt{\Big|{\frac{1}{3}}+u^A_f(t-\tau_h)\Big|}-\sqrt{{\frac{1}{3}}}\Bigg)\sin(k_j x_f)
\nonumber \\ &\quad + \dfrac{K_{\tau c}}{\gamma M} \Big(p^B_c(t-{\tau}_c)-p^A_c(t)\Big)\sin(k_j x_c),
\label{eq:gov2} 
\end{align}
\begin{align}
{{\dot{U}}_j}^B + \dfrac{k_j}{r}P_j^B = 0, 
\label{eq:gov3}
\end{align}
\begin{align}
{{\dot{P}}_j}^B &+ \dfrac{2\zeta_j\omega_j}{r}{P_j}^B -\dfrac{k_j}{r}U_j^B \nonumber\\
&= \dfrac{W^B}{r}\Bigg(\sqrt{\Big|\frac{1}{3} +u_f^B(t-\tau_h)\Big|} -\sqrt{\frac{1}{3}}\Bigg)\sin\left(\dfrac{k_jx_f}{r}\right)
\nonumber \\ 
&\quad + \dfrac{K_{\tau c}}{\gamma M r}\Big(p^A_c(t-{\tau}_c)-p^B_c(t)\Big)\sin\left(\dfrac{k_jx_c}{r}\right), 
\label{eq:gov4}
\end{align}
where $\omega_j=j\pi$ refers to the non-dimensional angular frequency of the $j$\textsuperscript{th} mode of oscillator A. $u_f(t-\tau_h) = u(x_f, t-\tau_h)$ and $p_c(x,t) = p(x_c, t)$. $W$ is the non-dimensional heater power given by:
\begin{align}
    W = \frac{4(\gamma - 1)L_w}{\gamma M c_0 p_0 S \sqrt{3}} (T_w - T_0) \sqrt{\pi \lambda_T C_v u_0 \rho_0 r_w}.
\end{align}
The frequency dependent damping, $\zeta_j$, in Eqs.~\eqref{eq:gov2} and \eqref{eq:gov4} is given by \cite{sterling1991nonlinear,matveev2003thermoacoustic}:
\begin{align}
     \zeta_j=\frac{1}{2\pi}\left(c_1\frac{\omega_j}{\omega_1}+c_2\sqrt{\frac{\omega_1}{\omega_j}}\right),  
\end{align}
where $c_1$ and $c_2$ are the damping coefficients. In the absence of coupling, varying the control parameters $W$, $c_1$, $c_2$, $x_f$, and $\tau_h$ in the model can result in the occurrence of limit cycle oscillations (LCOs) in a Rijke tube via subcritical Hopf bifurcation \cite{balasubramanian2008thermoacoustic,subramanian2010bifurcation}. All of the parameters in the model are non-dimensional, unless otherwise specified. Based on previous theoretical studies \cite{subramanian2010bifurcation,balasubramanian2008thermoacoustic,thomas2018effect}, we choose the values of the model parameters as in Table~\ref{tab:table1} for all the analytical approximations and numerical simulations in this study. We use ten Galerkin modes ($N=10$) in our simulations since we observe the coupled dynamics to remain the same on inclusion of higher modes \cite{thomas2018effect}. 
\begin{table}[b]
\caption{\label{tab:table1}%
Values of parameters kept constant for all numerical simulations of the model.
}
\begin{ruledtabular}
\begin{tabular}{cc|cc}
\textrm{Parameter}&
\textrm{Value}&
\textrm{Parameter}&
\textrm{Value}\\
\colrule
$\gamma$ & 1.4 & $l_A$ & 1 m\\
$M$ & $0.01$ & $x_f$ & $0.25$ \\
$c_1$ & 0.1 & $\tau_h$ & 0.2\\
$c_2$ & 0.06 & $x_c$ & 0.5\\
\end{tabular}
\end{ruledtabular}
\end{table}

\section{\label{sec:III}Results and discussions\protect}

\subsection{\label{sec:IIIA}Analysis of two delay coupled identical Rijke tube oscillators}

In this section, we numerically and analytically study the dynamical behavior of two coupled horizontal Rijke tube oscillators [see Fig.~\ref{fig:1}], and compare this behavior with that of an isolated oscillator. 

\begin{figure*}
\includegraphics[width=14cm]{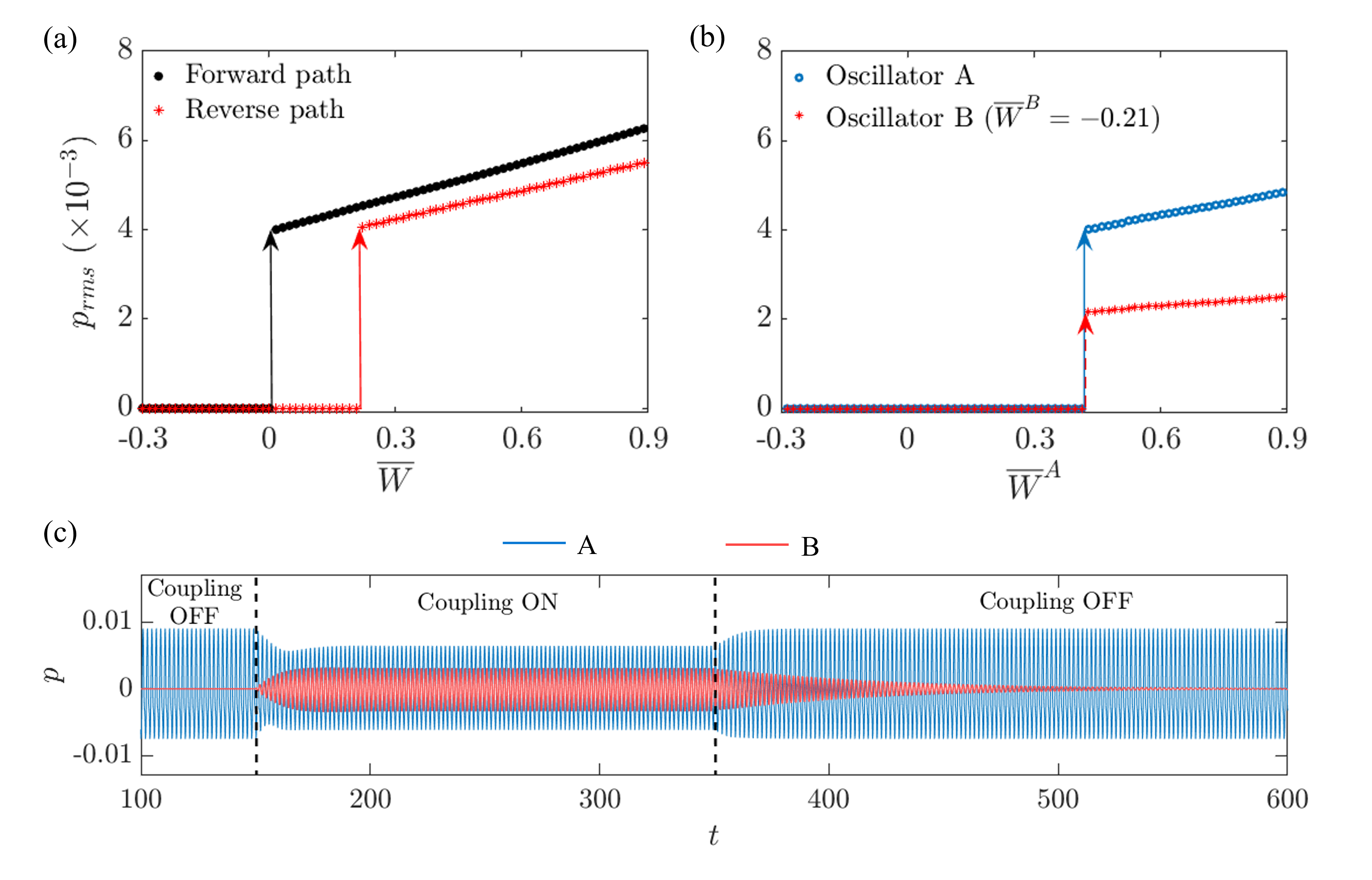}
\caption{\label{fig:2}One-parameter bifurcation diagrams between the root-mean-square value of acoustic pressure fluctuations, $p_{rms}$, and the normalized heater power, $\overline{W}$, (a) for an isolated oscillator A, and for two delay coupled oscillators A and B when their heater powers $W^A$ and $W^B$, respectively, are varied together (i.e., $\overline{W}^A = \overline{W}^B = \overline{W}$) and (b) for delay coupled oscillators A and B when $W^B$ is fixed at a low value ($\overline{W}^B=-0.21$) such that oscillator B is in steady state (i.e., $u^B=p^B=0$) before coupling, while $\overline{W}^A$ is varied until the system exhibits limit cycle oscillations. (c) Temporal variation of acoustic pressure oscillations, $p$, of two delay coupled oscillators shows the presence of induced oscillations on coupling an oscillator B that is initially in steady state ($\overline{W}^B=-0.21$) with another oscillator A exhibiting limit cycle oscillations ($\overline{W}^A=0.59$). $\tau_c=0.7$ and $K_{\tau c}=0.1$ are fixed for all the plots.}
\end{figure*}
We first consider the case when both the Rijke tubes are identical, i.e., $l_B=l_A=l$ and so $r=1$. In contrast to the study by Thomas \textit{et al.} \cite{thomas2018effect}, in the present study, we set unequal initial conditions in oscillators A and B to differentiate between the two identical oscillators. This, in turn, helps us obtain the distinct states of in-phase and anti-phase synchronization in the coupled Rijke tube oscillators, which is otherwise not possible to obtain. We now investigate how delay coupling influences the inherent bifurcations present in a Rijke tube oscillator.

\subsubsection{\label{sec:IIIA1}Comparison of bifurcations in isolated and delay coupled Rijke tube oscillators}

The one-parameter bifurcation diagrams in Fig.~\ref{fig:2}(a) illustrate the variation of the root-mean-square of the acoustic pressure signal ($p_{rms}$) on increasing the normalized heater power ($\overline{W}$) for two identical thermoacoustic oscillators when they are isolated and when they are delay coupled to each other. Here, we obtain the normalized heater power by normalizing $W$ by $W_H$, which is the critical value of heater power at the Hopf point of the isolated oscillator, i.e., $\overline{W} = W/W_H - 1$. Therefore, for an isolated oscillator, $\overline{W}=0$ at the Hopf point. 

In Fig.~\ref{fig:2}(a), the normalized heater powers, $\overline{W}^A$ and $\overline{W}^B$, of the coupled identical Rijke tube oscillators are varied together (i.e., $\overline{W}^A = \overline{W}^B = \overline{W}$). On increasing the value of $\overline{W}$, we observe that both the isolated and the delay coupled identical oscillators undergo subcritical Hopf bifurcation, wherein the oscillators transition abruptly from a state of stable fixed point to limit cycle oscillations. However, we find that the Hopf point of the oscillators when they are coupled to each other is higher ($\overline{W}=0.22$) than that of the isolated oscillators ($\overline{W}=0$). Increasing the value of $\overline{W}$ beyond the Hopf point of the isolated or the coupled oscillators leads to a corresponding growth in the amplitude of the limit cycle oscillations [refer to Fig.~\ref{fig:2}(a)]. We find the limit cycle oscillations in the coupled system to be smaller in amplitude as compared to the uncoupled oscillator. In the present study, we restrict our analysis to when the Rijke tubes individually exhibit period-1 limit cycle oscillations with the first mode being dominant. Accordingly, we vary the value of $\overline{W}$ upto 0.9 since we observe period-2 oscillations in the isolated Rijke tube oscillator for $\overline{W}>0.9$.


In another case shown in Fig.~\ref{fig:2}(b), we consider the same system of two delay coupled oscillators. However, here we vary only the normalized heater power of oscillator A ($\overline{W}^A$) while the normalized heater power of oscillator B ($\overline{W}^B$) is fixed at a low value ($\overline{W}^B=-0.21$) so that it is in steady state prior to coupling. We observe that the shift in the Hopf point of oscillator A is greater ($\overline{W}^A = 0.42$) as compared to the case when the normalized heater powers of both the oscillators are equally varied [Hopf point is at $\overline{W}=0.22$ in Fig.~\ref{fig:2}(a)]. We further notice that when oscillator A exhibits LCOs, it induces small amplitude periodic oscillations in oscillator B, even though oscillator B is always in steady state prior to coupling. Figure \ref{fig:2}(c) illustrates these coupling-induced periodic oscillations in oscillator B for $\overline{W}^A=0.59$ and $\overline{W}^B=-0.21$. Since both the oscillators have similar natural frequencies, on coupling [at $t=150$ in Fig.~\ref{fig:2}(c)], oscillator A drives oscillator B close to its resonant frequency. As a result, we observe induced periodic oscillations of low amplitude in oscillator B, which can also be seen from the bifurcation plot in Fig.~\ref{fig:2}(b). On removing the coupling [at $t=350$ in Fig.~\ref{fig:2}(c)], the oscillations in oscillator B die down, while those in oscillator A regain their original amplitude observed in the uncoupled state. This further shows that in a system of two delay coupled Rijke tubes, when one Rijke tube oscillator is in the state of limit cycle oscillations, another oscillator can never be in steady state due to coupling-induced low amplitude periodic oscillations. Note that Figs.~\ref{fig:2}(a) and (b) illustrate only the effect of increasing the heater power on the Rijke tubes in the forward path (the reverse path is not shown) for clarity.

\begin{figure*}
\includegraphics[width=13cm]{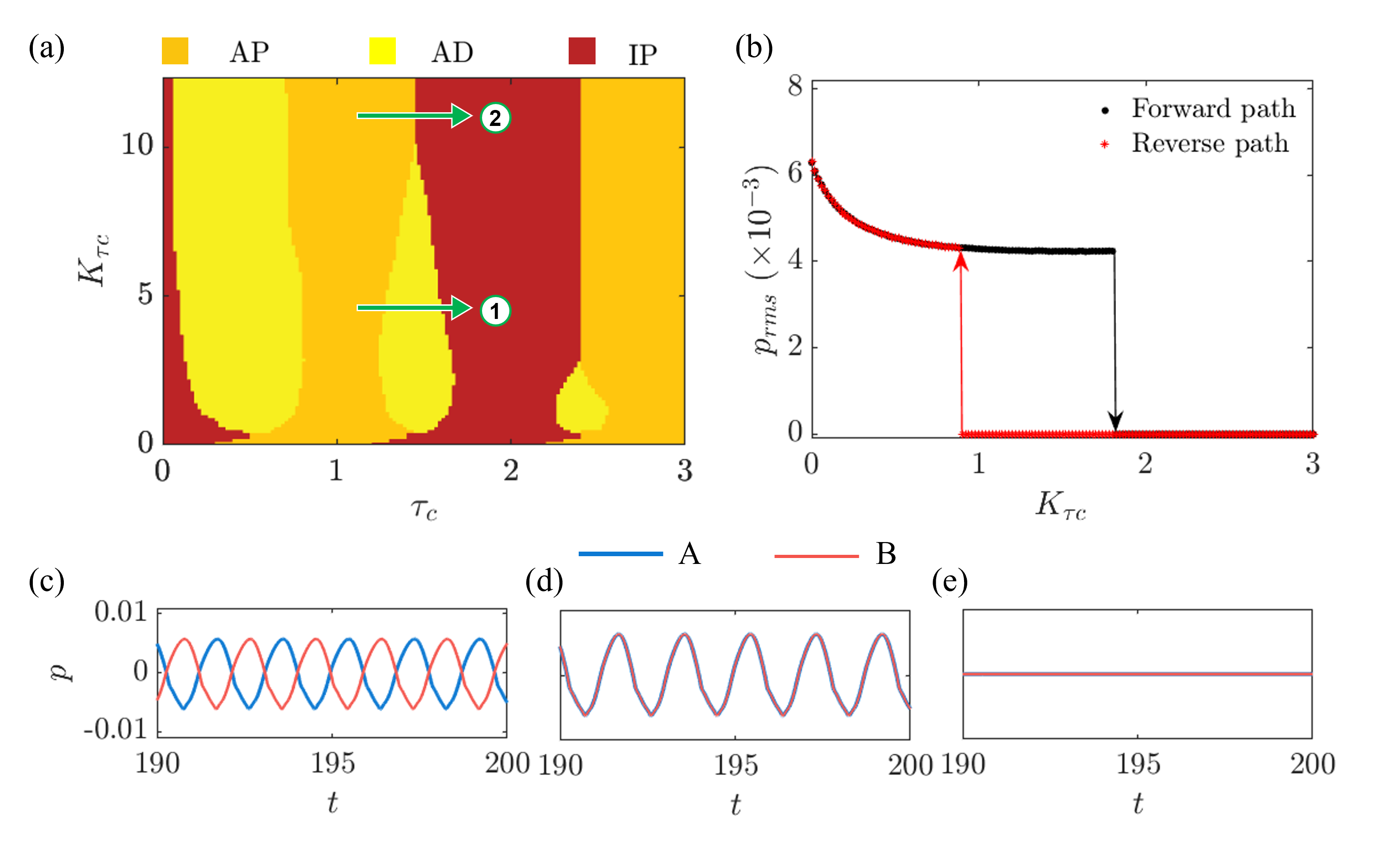}
\caption{\label{fig:3}Two-parameter bifurcation diagram between mutual delay coupling strength, $K_{\tau c}$, and coupling delay, $\tau_c$, for two delay coupled identical Rijke tube oscillators depicts the occurrence of multiple islands of amplitude death (AD) in the system. The arrows indicate the two distinct routes through which the system transitions between the states of in-phase synchronization from anti-phase synchronization: The first route (labelled `1') is via an intermediate state of AD, while the second route (labelled `2') is through phase-flip bifurcation. (b) One-parameter bifurcation diagram showing the variation of the root-mean-square value of the non-dimensional acoustic pressure ($p_{rms}$) with $K_{\tau c}$ for $\tau_c=0.7$ illustrates explosive hysteretic transition to AD. Non-dimensional acoustic pressure signals ($p$) of both the oscillators corresponding to the states of (c) in-phase synchronization (IP), (d) amplitude death (AD), and (e) anti-phase synchronization (AP). The value of $\overline{W}$ is fixed at 0.59 for both the oscillators.}
\end{figure*}

\subsubsection{\label{sec:IIIA2}Amplitude death, phase-flip bifurcation, and hysteresis in delay coupled identical Rijke tube oscillators}

In this subsection, we systematically study the effect of the following parameters on the interaction of delay coupled identical Rijke tube oscillators: (i) the normalized heater power ($\overline{W}$), which is a system parameter that directly affects the amplitude of limit cycle oscillations in the uncoupled state [as seen in Fig.~\ref{fig:2}(a)], and coupling parameters such as (ii) the delay coupling strength ($K_{\tau c}$) and (iii) the coupling delay ($\tau_c$). For each combination of parameter values, we first let both the Rijke tubes exhibit limit cycle oscillations in isolation and after that we initiate the coupling between them to study their behavior. We quantify the suppression of the acoustic pressure oscillations due to coupling as $\Delta p=p_{rms,0}-p_{rms}$, where $p_{rms,0}$ and $p_{rms}$ are the root-mean-square values of the limit cycle oscillations before and after coupling, respectively. This value of $\Delta p$ is then normalized with respect to $p_{rms,0}$ to get the relative amplitude suppression ($\Delta p/p_{rms,0}$). When $\Delta p/p_{rms,0}=1$, the oscillations are completely quenched, while $\Delta p/p_{rms,0}=0$ corresponds to the absence of any suppression in the limit cycle oscillations on coupling.

In Fig.~\ref{fig:3}(a), we show the two-parameter bifurcation diagram between $K_{\tau c}$ and $\tau_c$, which illustrates the effect of varying coupling parameters in two time-delay coupled Rijke tubes exhibiting high amplitude limit cycle oscillations in their uncoupled state [$\overline{W}=0.59$, refer to Fig.~\ref{fig:2}(a)]. We observe three distinct states in the system, which are classified as in-phase synchronization (IP) [Fig.~\ref{fig:3}(c)], anti-phase synchronization (AP) [Fig.~\ref{fig:3}(e)], and amplitude death (AD) [Fig.~\ref{fig:3}(d)]. The system is said to be in the state of in-phase synchronization (IP) when the phase difference between the acoustic pressure oscillations in oscillators A and B is close to 0 deg. On the other hand, the phase difference is nearly 180 deg when the oscillators are in the state of anti-phase synchronization (AP). Analytic signal approach based on Hilbert transform is utilized to extract the instantaneous phase of the acoustic pressure signals during the state of limit cycle oscillations in the two Rijke tubes \cite{pikovsky2003synchronization}. When the coupled limit cycle oscillations are synchronized, the relative phase between them fluctuates in time around a constant value. This constant phase difference ($\overline{|\Delta \phi|}$) is calculated as the arithmetic mean of the absolute difference between the instantaneous phases of the oscillations in oscillators A and B. During amplitude death (AD), both the Rijke tube oscillators approach the same steady state (or fixed point) upon coupling. The regions of AD in the bifurcation diagram [Fig.~\ref{fig:3}(a)] are present around values of coupling delay $\tau_c =$ 1/2, 3/2, 5/2, …, which approximately correspond to odd-multiples of quarter-period of the oscillations, such as $T/4$, $3T/4$, $5T/4$,…, where $T \approx 2l/c_0$ is the time period of the limit cycle oscillations in the absence of coupling.
\begin{figure*}
\includegraphics[width=14cm]{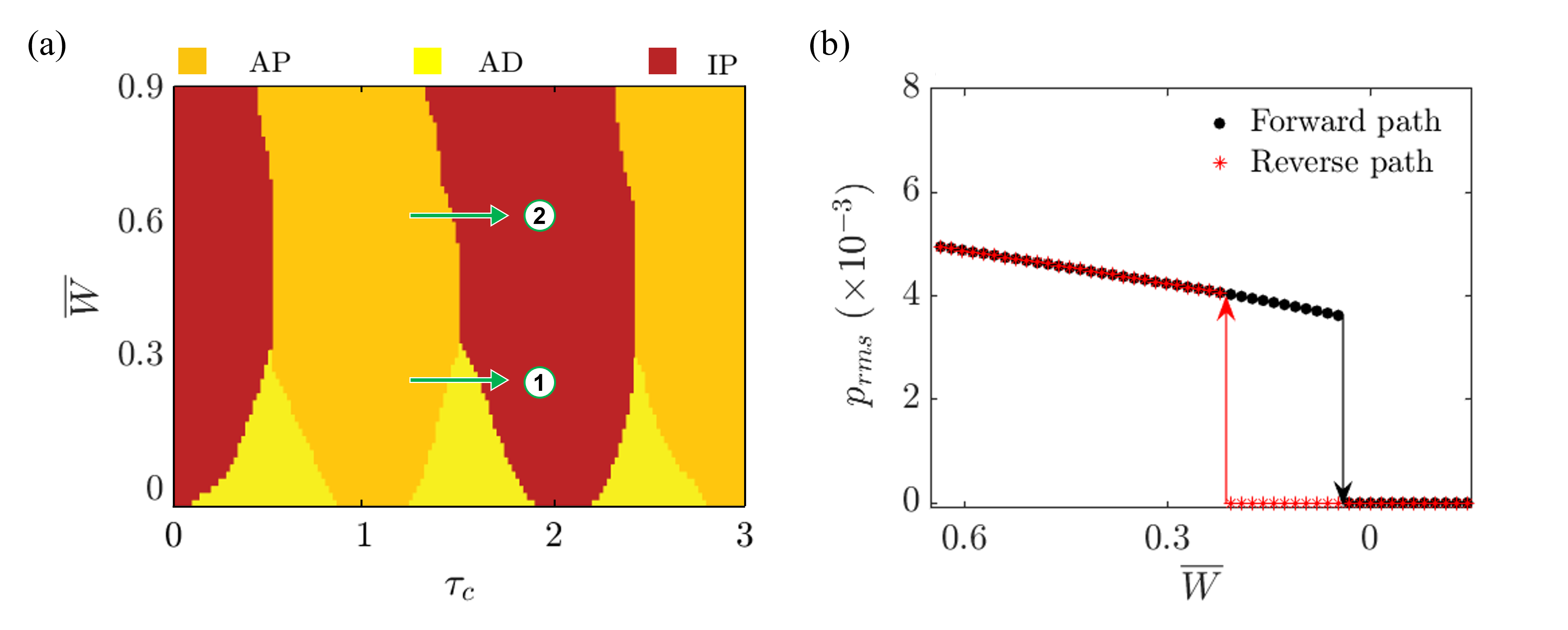}
\caption{\label{fig:4}(a) Two-parameter bifurcation diagram between the normalized heater power ($\overline{W}$) and the coupling delay ($\tau_c$) for two time-delay coupled identical Rijke tube oscillators. The arrows indicate the two routes through which the system transitions between in-phase synchronization and anti-phase synchronization, as discussed in Fig.~\ref{fig:3}. (b) One-parameter bifurcation plot between the root-mean-square value of acoustic pressure oscillations, $p_{rms}$, and $\overline{W}$ for $\tau_c=0.7$ depicts explosive hysteretic transition to AD. $K_{\tau c}$ is fixed at 0.1 in both the plots.}
\end{figure*}

In Fig.~\ref{fig:3}(a), the regions of AD manifest as islands, except for the first AD region which does not disappear on increasing the coupling strength to very high values. These AD regions decrease in size with increase in $\tau_c$. The AD regions are surrounded by regions of IP and AP synchronization, which occur alternately with increasing $\tau_c$ [depicted by the arrow `1' in Fig.~\ref{fig:3}(a)]. Away from the AD regions, the system undergoes abrupt transitions from IP to AP state or vice-versa on varying $\tau_c$ [depicted by the arrow `2' in Fig.~\ref{fig:3}(a)]. Such an abrupt transition in the phase difference between the oscillators is referred to as phase-flip bifurcation (PFB) \cite{prasad2006phase}. 

Now, we study how the coupled system transitions from the state of LCOs to AD or vice-versa on varying $K_{\tau c}$ using a one-parameter bifurcation diagram between the root-mean-square of the acoustic pressure signal ($p_{rms}$) and $K_{\tau c}$ for a constant $\tau_c = 0.7$ [depicted in Fig.~\ref{fig:3}(b)]. We observe that the coupled system undergoes fold bifurcation during the transition from LCO to AD state (in the forward path) and subcritical Hopf bifurcation when transitioning from AD to LCO state (in the reverse path), resulting in hysteresis on variation of $K_{\tau c}$. Thus, coupled Rijke tube oscillators exhibit `explosive' (first-order) transition \cite{kuehn2021universal} during the occurrence of amplitude death.

Recent studies have demonstrated that apart from coupling parameters, system parameters such as the amplitude and the natural frequency of an oscillator also play a significant role in determining the behavior of mutually coupled oscillators \cite{dange2019oscillation,premraj2021effect}. Therefore, we next investigate the effect of system parameters on the coupled behavior of Rijke tube oscillators. Towards this purpose, we vary the amplitude of the acoustic pressure oscillations in the uncoupled state by varying the heater powers ($W^{A,B}$) in Eqs.~\eqref{eq:gov3} and \eqref{eq:gov4} equally for both the oscillators. Note that the oscillators exhibit LCOs before the initiation of coupling. 

Figure~\ref{fig:4}(a) depicts the two-parameter bifurcation diagram between normalized heater power ($\overline{W}$) and coupling delay ($\tau_c$) for two time-delay coupled identical Rijke tube oscillators. For lower values of $\overline{W}$, we notice that the dynamics of the system alternates between in-phase synchronization (IP) and anti-phase synchronization (AP) via an intermediate state of amplitude death (AD) as $\tau_c$ is increased [shown by the arrow `1' in Fig.~\ref{fig:4}(a)]. When $\overline{W}$ is sufficiently high, the coupled behavior of the oscillators switches abruptly between IP and AP states by undergoing PFB [shown by the arrow `2' in Fig.~\ref{fig:4}(a)]. 

The one-parameter bifurcation diagram in Fig.~\ref{fig:4}(b) illustrates the variation in the root-mean-square value of the acoustic pressure oscillations ($p_{rms}$) with $\overline{W}$ for a fixed value of $\tau_c$. We find that the coupled Rijke tube oscillators undergo fold bifurcation while transitioning from LCO to AD state when decreasing $\overline{W}$ in the forward path. On the other hand, the coupled oscillators undergo subcritical Hopf bifurcation during the transition from AD to LCO state at a higher value of $\overline{W}$ in the reverse path. Thus, the system undergoes explosive hysteretic transitions between AD and LCO states on variation of system parameters [refer to Fig.~\ref{fig:4}(b)] and also coupling parameters [refer to Fig.~\ref{fig:3}(b)] when the Hopf point of the individual oscillators are subcritical in nature. 

In the present model, the Rijke tube oscillators can only exhibit subcritical Hopf bifurcation due to the nature of the nonlinearity in the heat release rate fluctuations shown in Eq.~\eqref{eq:heckl} \cite{subramanian2010bifurcation}. However, experiments by Etikyala and Sujith \cite{etikyala2017change} show that Rijke tubes can undergo supercritical Hopf bifurcation for low air flow rates. Hence, in Sec.~I of the Supplementary Material, we modify the nonlinear terms in the model of delay coupled thermoacoustic oscillators so that the oscillators exhibit supercritical Hopf bifurcation in isolation \cite{tandon2020bursting}. We observe that delay coupled thermoacoustic oscillators with supercritical Hopf points exhibit second-order (i.e., continuous) change in the amplitude of acoustic pressure fluctuations during the transition from LCO to AD state and vice-versa, without hysteresis. Thus, we confirm that the nature of the route to AD on variation of system and coupling parameters for coupled Rijke tube oscillators depends on the criticality of the bifurcation exhibited by the constituent oscillators in the uncoupled state. 

\begin{figure}
\includegraphics[width=8.4cm]{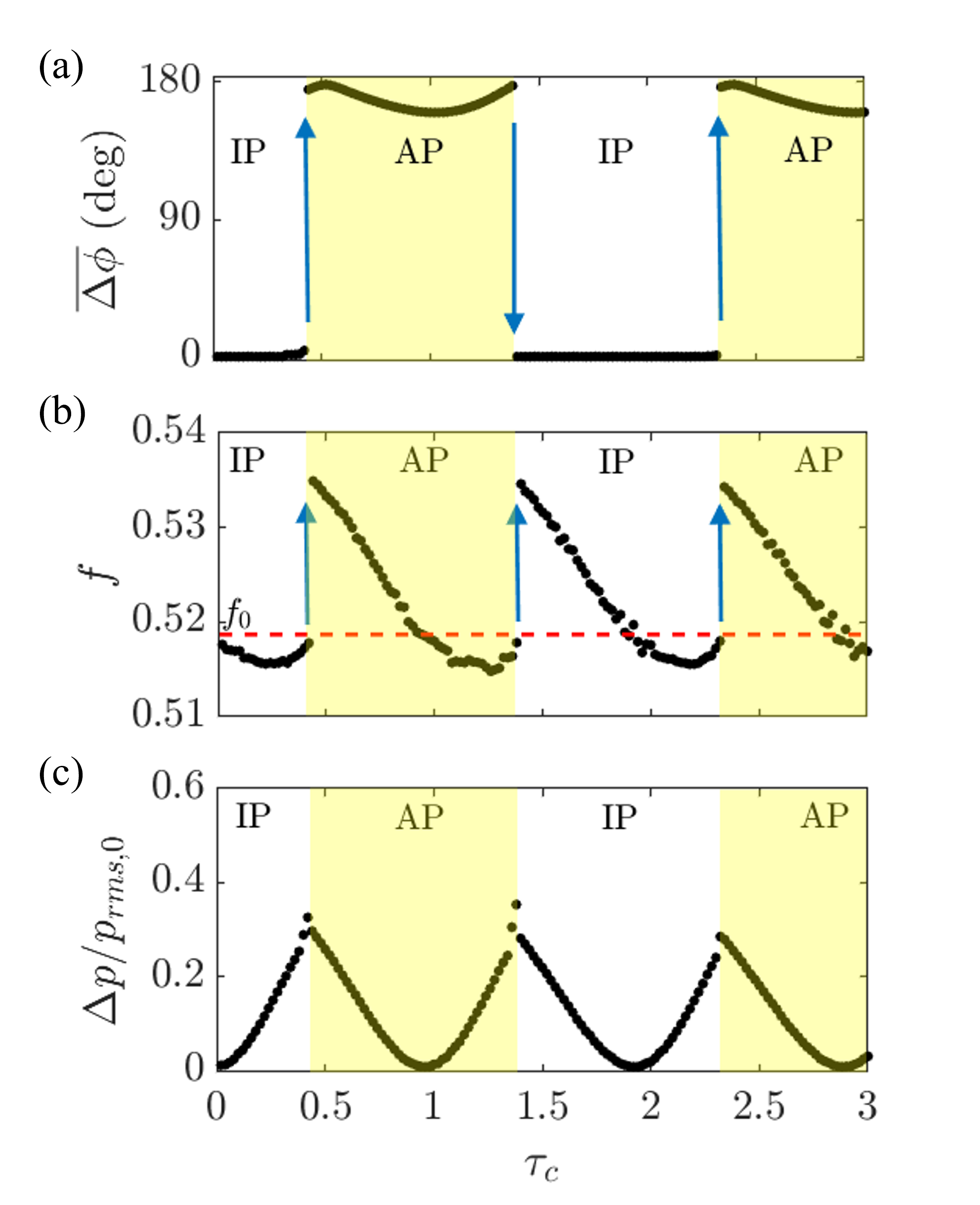}
\caption{\label{fig:5}Occurrence of phase-flip bifurcation (PFB) on variation of coupling delay ($\tau_c$) in two delay coupled identical Rijke tube oscillators. Variation in (a) the mean phase difference ($\overline{|\Delta \phi|}$) between the acoustic pressure oscillations in the coupled Rijke tube oscillators A and B, and (b) the non-dimensional dominant frequency ($f$) of the synchronized oscillations in the system, and (c) relative amplitude suppression ($\Delta p/p_{rms,0}$) as a function of $\tau_c$ for $\overline{W} = 0.59$. The non-dimensional frequency of the oscillations of the uncoupled system ($f_0$) is shown by the red dashed line in (b), whose dimensional value is 176 Hz. $K_{\tau c}=0.1$ is fixed for all plots. Discontinuous changes in the values of ($\overline{|\Delta \phi|}$) and $f$ are denoted by arrows.}
\end{figure}

Next, we take a closer look at how the properties of the acoustic pressure oscillations change as the system undergoes phase-flip bifurcations. Towards this purpose, we plot the variation of the mean phase difference ($\overline{|\Delta \phi|}$) between the LCOs of the two Rijke tubes [see Fig.~\ref{fig:5}(a)], the non-dimensional dominant frequency ($f$) of the synchronized oscillations [see Fig.~\ref{fig:5}(b)], and the relative suppression ($\Delta p/p_{rms,0}$) in the amplitude of the oscillations [see Fig.~\ref{fig:5}(c)] as a function of the coupling delay $\tau_c$. The heater power is set high enough ($\overline{W} = 0.59$) so that the system always exhibits LCOs for all values of $\tau_c$ for $K_{\tau c}=0.1$ [see Fig.~\ref{fig:4}(a)]. On varying the coupling delay $\tau_c$ [in Figs.~\ref{fig:5}(a)], we notice a sudden change in the value of ($\overline{|\Delta \phi|}$) from near 0 deg to 180 deg and vice-versa when the system undergoes PFB. The non-dimensional dominant frequency of the system ($f$) exhibits an oscillatory behavior on varying $\tau_c$ [Fig.~\ref{fig:5}(b)]. It jumps whenever the system transitions from IP to AP state or vice-versa. Post the jump, we notice that the value of $f$ falls almost linearly until it crosses the value of the natural frequency [$f_0$, shown by the dashed line in Fig.~\ref{fig:5}(b)]. The frequency then varies nonlinearly with $\tau_c$ until it again approaches $f_0$, after which it jumps once again at the next bifurcation point. From Fig.~\ref{fig:5}(c), we observe that the extent of suppression of limit cycle oscillations displays an oscillatory behavior where the amplitude suppression increases as the system approaches the point of PFB and decreases post PFB. We do not observe hysteresis in the dynamical properties of the LCOs in the coupled Rijke tube system around PFB. 

In Sec.~II of the Supplementary Material, we compare the results from the model with the corresponding experimental results obtained by Dange \textit{et al} \cite{dange2019oscillation}. We first compare a portion of Figs.~\ref{fig:4}(a), \ref{fig:5}(a), and \ref{fig:5}(b) with the corresponding experimental results in Fig. S2. We also show the comparison of the trends in amplitude suppression for different values of coupling strength in Fig.~S3. We observe qualitative similarity between the results from the model and the experiments on coupled Rijke tube oscillators.

\subsubsection{\label{sec:IIIA3} Analytical approximation for delay coupled identical Rijke tube oscillators}

Through numerical simulations, we have so far determined the effect of system parameters and coupling parameters on the behavior of delay coupled identical Rijke tube oscillators. We observed the occurrence of AD and PFB in the coupled Rijke tube system. We will now attempt to explain how these phenomena occur in the system by analytically deriving an approximate solution of the model.

We start our analysis by combining the governing equations for identical Rijke tube oscillators [Eqs.~\eqref{eq:gov1}-\eqref{eq:gov4} with $r=1$ and $W^A = W^B = W$] into a single set of second-order delay differential equations in terms of $U_j^{A,B}$ [since the acoustic pressure $P_j^{A,B}$ can be expressed in terms of $\dot{U}_j^{A,B}$ using Eq.~\eqref{eq:gov1} and \eqref{eq:gov3}]. Note that subscripts are used for denoting the oscillators instead of superscripts for convenience. The frequency dependent damping in the present model preferentially damps higher modes \cite{subramanian2013subcritical}. Thus, since the first mode is dominant in our system, we consider only $j=1$ and neglect the effect of higher modes to simplify the equations \cite{subramanian2013subcritical}. We subsequently drop the subscript $j$ to yield the following delay differential equations for the system of two delay coupled Rijke tubes: 
\begin{align}
\label{eq:7}
    \ddot{U}_{A,B} &+ 2\zeta \omega \dot{U}_{A,B}+k^2 U_{A,B} + Wk\sin(kx_f)
    \nonumber\\ 
    & \times \Big[\sqrt{|1/3+\cos(kx_f) U_{A,B} (t-\tau_h)|}-\sqrt{1/3}\Big]\nonumber\\ 
    &  \qquad = K_{\tau c}\sin^2(kx_c) (\dot{U}_{B,A} (t-\tau_c) - \dot{U}_{A,B}). 
\end{align}
To determine the stability of the steady state in the system, we locate the parameter values where the trivial solution for acoustic velocity (i.e., $U=0$) loses its stability. We therefore assume small amplitudes for $U$ and linearize the square-root nonlinearity in Eq.~\eqref{eq:7} using Taylor series expansion as we are only interested in how the system behaves close to the steady state \cite{subramanian2013subcritical}. The resulting equation is: 
\begin{align}
\label{eq:8}
     \ddot{U}_{A,B} &+b_0 \dot{U}_{A,B}  + b_1U_{A,B} +  \sigma U_{A,B}(t-\tau_h) \nonumber\\
     &+ K_{\tau c}\sin^2(kx_c)(\dot{U}_{A,B}-\dot{U}_{B,A} (t-\tau_c))=0, 
\end{align} 
where $b_0=2 \zeta \omega=2\pi \zeta$, $\sigma=(\sqrt{3}/4)Wk \sin(2kx_f)$, and $b_1=k^2=\pi^2$. 

In order to further simplify the above equation, we employ the method of averaging \cite{balanov2008synchronization,wahi2005study}, for which we assume small values for $W$ and $K_{\tau c}$, $\tau_c$, and $\tau_h$ so that the assumption of slowly varying amplitudes holds true during the method of averaging. Then, by considering a symmetric solution (i.e., identical oscillators exhibit oscillations of the same amplitude), the method of averaging on Eq.~\eqref{eq:8} yields the following slow flow equations for the amplitude ($R$) and the phase ($\phi_A$ and $\phi_B$ for oscillator A and B, respectively) of the oscillations of the coupled system (refer to Supplementary Material Sec.~III for the complete derivation):  
\begin{equation}
\label{eq:9}
\dot{R}=\frac{R}{2}\Big[-K_{\tau c}\sin^2(kx_c)(1-|\cos(\omega \tau_c)|) +\frac{\sigma}{\omega} \sin(\omega \tau_h)-b_0 \Big],
\end{equation}
\begin{align}
\label{eq:10}
\dot{\phi}_{A,B} &= -\frac{\omega}{2} + \frac{b_1}{2\omega} + \frac{\sigma}{2\omega} \cos(\omega \tau_h) \nonumber\\
&\quad -\frac{K_{\tau c}\sin^2(kx_c)}{2}\sin[\omega \tau_c - (\phi_{B,A}-\phi_{A,B})].
\end{align}
From Eq.~\eqref{eq:10}, by subtracting the equation for $\dot{\phi}_{A}$ from that of $\dot{\phi}_{B}$, we get the slow flow equation for the phase difference between the oscillators ($\theta=\phi_B-\phi_A$) as follows:
\begin{align}
\label{eq:11}
\dot{\theta}=-K_{\tau c}\sin^2(kx_c) \cos(\omega \tau_c) \sin(\theta).   
\end{align}

\begin{figure}[t]
\includegraphics[width=6.8cm]{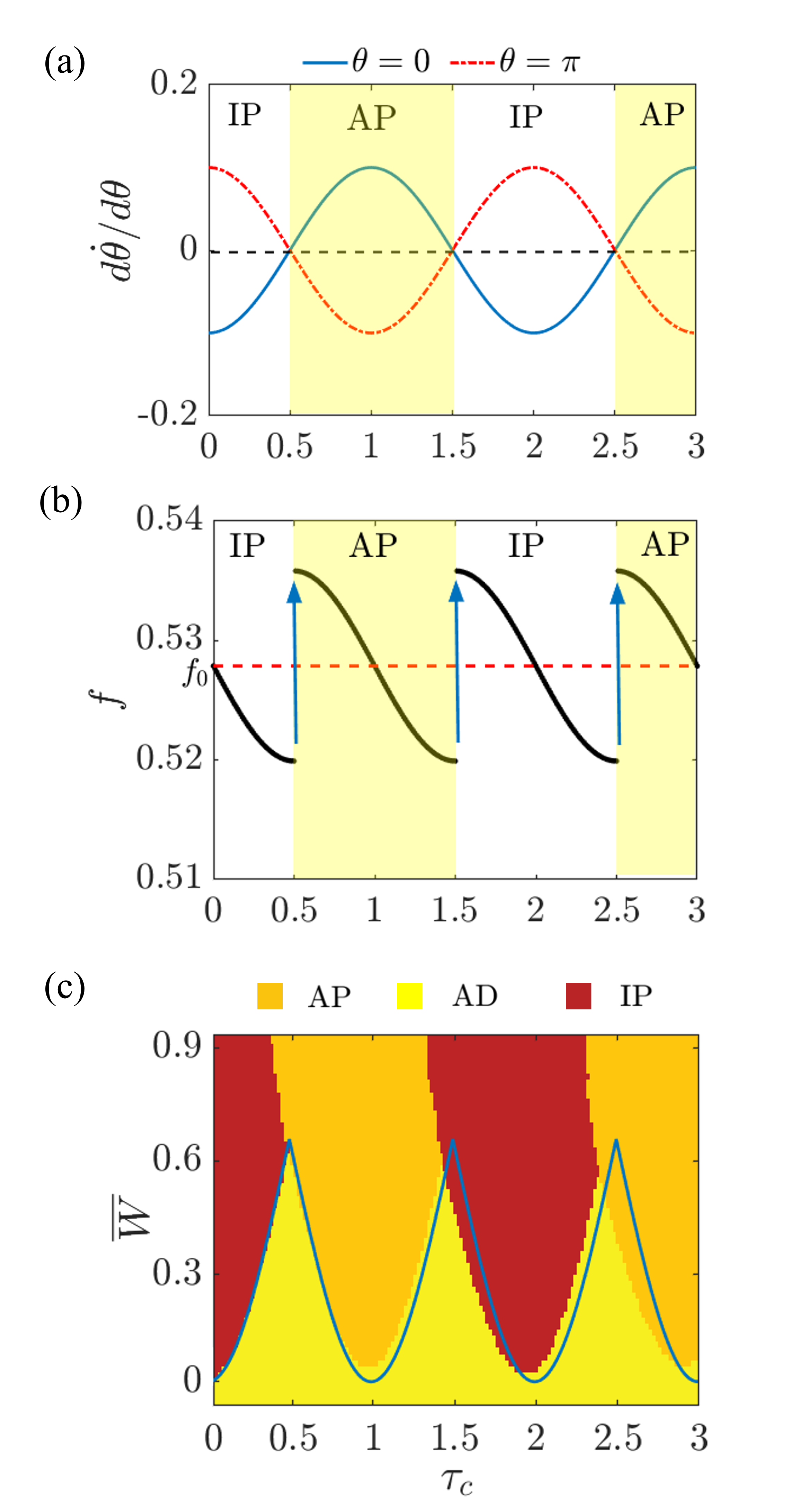}
\caption{\label{fig:6} The occurrence of AD and PFB in two mutually delay coupled Rijke tube oscillators determined using the method of averaging. (a) Variation in the stability of the states of in-phase synchronization ($\theta=0$ rad) and anti-phase synchronization ($\theta=\pi$ rad) on varying the mutual coupling delay ($\tau_c$). PFB occurs when the sign of $d\dot{\theta}/d\theta$ changes, i.e., when the curves cross the horizontal dashed line. (b) Variation in the frequency of the oscillations ($f$) in the coupled system on varying $\tau_c$. The frequency of the uncoupled oscillations ($f_0$) is illustrated by the horizontal dashed line. (c) The two-parameter bifurcation plot between $\overline{W}$ and $\tau_c$ obtained numerically is overlaid with the analytically obtained boundary (blue line) demarcating the AD region. $K_{\tau c}=0.1$ is fixed for all plots.}
\end{figure}

Let us first consider the above equation [Eq.~\eqref{eq:11}] for the phase difference, $\theta$. We observe the presence of two principal values of fixed points by setting the time derivative as zero: 0 and $\pi$ rad, assuming $\cos(\omega \tau_c)\neq 0$. This indicates that in-phase synchronized oscillations ($\theta=0$) and anti-phase synchronized oscillations ($\theta=\pi$) are the only possible steady state solutions of the system. The stability of these fixed points, 0 and $\pi$, can be determined by examining the sign of the derivative \cite{strogatz2018nonlinear} $d\dot{\theta}/ d\theta=-K_{\tau c}\sin^2(kx_c) \cos(\omega \tau_c) \cos(\theta)$ for $\theta=0$ and $\theta=\pi$, plotted in Fig.~\ref{fig:6}(a) as a function of $\tau_c$. We notice that each of the fixed points become alternately stable and unstable on varying the mutual coupling delay. At $\tau_c=  1/2$, $3/2$, $5/2$, …, the signs of the derivative changes, indicating that the in-phase synchronized solution loses stability while the anti-phase synchronized state becomes stable, or vice-versa. This implies the periodic occurrence of phase-flip bifurcation (PFB), i.e., the abrupt transition between IP and AP states, on varying the coupling delay [as already seen in Figs.~\ref{fig:3}(a), \ref{fig:4}(a), and \ref{fig:5}(a)].

To understand how PFB gives rise to frequency jumps in the system, we consider the slow flow equation for the phase of each oscillator [Eq.~\eqref{eq:10}]. The phase of the oscillator is $\omega t + \phi$ and so its angular frequency is $\omega + \dot{\phi}$. Hence, the frequency of the oscillations of the mutually delay coupled system is given by  $f=1/2\pi (\omega + \dot{\phi})=1/2+(1/2\pi) \dot{\phi}$, since $\omega = \pi$ for the first mode. Note that the right-hand side of Eq.~\eqref{eq:10} has the term $\phi_{B,A}-\phi_{A,B}$, which is evaluated according to whether the IP state or the AP state is stable for that particular value of coupling delay $\tau_c$ [as shown in Fig.~\ref{fig:6}(a)]. At $\tau_c=0.5$, we notice that $\sin(\omega \tau_c)=1$. Around this value of coupling delay, the frequency of the oscillations is minimum for $\theta=0$, while the frequency is maximum for $\theta=\pi$. Hence, while undergoing PFB, the frequency of the oscillations also abruptly increases. A similar argument can be made for $\tau_c=3/2, 5/2,$ …

We also observe from Eq.~\eqref{eq:10} that for a particular value of the phase difference, the frequency varies sinusoidally with coupling delay. Hence, after the frequency jumps to a maximum value during PFB, it falls like a half-sine wave as the coupling delay is increased further. Once the value of the frequency reaches its minimum, PFB occurs once again. In this manner, the frequency of the oscillations of the delay coupled system undergoes periodic variation with coupling delay. Figure \ref{fig:6}(b) illustrates the resulting trends in the frequency, $f$, of the oscillations with coupling delay for $K_{\tau c}=0.1$. We notice the analytically obtained frequency trends to match well with the corresponding numerical results in Fig.~\ref{fig:5}(b). We have thus modeled the occurrence of PFB in the system and investigated the underlying mechanism through our analytical approximations.

To uncover the conditions for amplitude death (AD) in the delay coupled identical Rijke tube oscillators, we look into the slow flow amplitude equation [Eq.~\eqref{eq:9}]. We notice that $R=0$ (which corresponds to AD state) is the only solution since we linearized the equations of the model. This solution loses stability when there is a change in the sign of the coefficient of $R$. Hence, the transition from AD to LCO occurs when $K_{\tau c}\sin^2(kx_c)(|\cos(\omega \tau_c|-1)+(\sigma/\omega) \sin(\omega \tau_h)-b_0=0$, which simplifies to the following:
\begin{equation}
\label{eq:12}
|\cos(\omega \tau_c)|<\Big[1-\Big(\frac{\sigma}{\omega} \sin(\omega \tau_h)-b_0\Big)/K_{\tau c}\sin^2(kx_c) \Big]. \end{equation}
The above equation succinctly illustrates how the interplay between system and coupling parameters determines the dynamical behavior of two delay coupled identical Rijke tube oscillators. 

Next, we qualitatively examine the role played by each of the parameters on the stability of the AD state. We fix the value of $\omega$ at $\pi$ (which is the non-dimensional angular frequency of the first mode) and consider a constant value of $\tau_h$ and $x_c$. We vary the coupling parameters $K_{\tau c}$ and $\tau_c$. The system parameters involved in Eq.~\eqref{eq:12} are $\sigma$ (which can be varied by varying the heater power, $W$), and $b_0$ (which is indicative of the damping in the model). AD is possible for a wider range of parameters when the values of right-hand-side and left-hand-side of Eq.~\eqref{eq:12} are high and low, respectively. This is possible for high coupling strength $K_{\tau c}$, small values of  $\sigma$, i.e., small values of heater power $W$ and high damping coefficients (which would give large value of $b_0$). On the other hand, the value of $|\cos(\omega \tau_c)|$ is least when $\cos(\omega \tau_c)=0$, i.e., when $\tau_c=1/2,3/2,5/2,$…. These are the optimal values of coupling delay for achieving AD in the system.  

In order to unravel the nature of the bifurcation between AD and LCO states in the coupled identical Rijke tubes, we include the cubic term in the Taylor series expansion of the square-root nonlinearity in Eq.~\eqref{eq:7}; this gives the equation for oscillator B as:
\begin{align}
\label{eq:13}
\ddot{U}_B + b_0 \dot{U}_B + b_1 U_B + \sigma U_B (t-\tau_h)+\sigma_2[U_B (t-\tau_h)]^2 \nonumber\\ +\sigma_3[U_B (t-\tau_h)]^3 + K_{\tau c}\sin^2(kx_c) \left[\dot{U}_B- \dot{U}_A (t-\tau_c)\right]=0,     
\end{align}
where $\sigma_2=-\frac{3}{4} \sigma  \cos(k x_f)$ and $\sigma_3=\frac{9}{8} \sigma \cos^2 (k x_f)$.
The equation for oscillator A is obtained by interchanging A and B in the above equation. By using the method of averaging (the steps are detailed in the Supplementary Material Sec.~III), with the same assumptions as before, we get the amplitude equation:
\begin{align}
\label{eq:14}
\dot{R}=\frac{R}{2}\Big[\frac{\sigma}{\omega} \sin(\omega \tau_h)-b_0 - K_{\tau c}\sin^2(kx_c)(1 +|\cos(\omega \tau_c)|) \Big]
\nonumber\\
+\frac{3}{8\omega}\sigma_3 \sin(\omega \tau_h)R^3.
\end{align}
The equation for the phase difference is the same as before [Eq.~\eqref{eq:11}]. The amplitude equation is of the form:
\begin{align}
\label{eq:15}
 \dot{R}=C_1 (\sigma-\sigma_H)R+C_2 R^3, 
\end{align}
where $C_1=\frac{1}{2\omega} \sin(\omega \tau_h)$, $\sigma_H=(\omega/\sin(\omega \tau_h)) \big[b_0+K_{\tau c}\sin^2(kx_c) (1 - |\cos(\omega \tau_c)|)\big]$, and $C_2= \frac{3\sigma_3}{8\omega} \sin(\omega \tau_h)$.
This equation is similar to the amplitude equation of the Stuart-Landau oscillator, which is the normal form of Hopf bifurcation \cite{provansal1987benard,subramanian2013subcritical}. Hence, the delay coupled system undergoes Hopf bifurcation at $\sigma=\sigma_H$.

The criticality of the Hopf bifurcation is determined by the sign of the coefficient of the cubic term in Eq.~\eqref{eq:15}. In our model, the coefficient $C_2$ is always positive and it is not dependent on coupling parameters. This indicates that the bifurcation is subcritical irrespective of whether the oscillators are coupled or not. Thus, the expression given by Eq.~\eqref{eq:12} predicts the set of subcritical Hopf points of the delay coupled system. We juxtapose the analytically predicted Hopf points with the corresponding numerically obtained bifurcation diagram in Fig.~\ref{fig:6}(c). We set small initial conditions and couple the oscillators at the start (before they individually reach the LCO state) so as to obtain the Hopf points, and not the fold points, in the numerical result in Fig.~\ref{fig:6}(c). We observe an excellent match between the analytical and the numerical results when our simplifying assumption of small coupling delay holds true. Recently, Premraj \textit{et al}. \cite{premraj2021effect} showed that delay coupled Stuart-Landau oscillators qualitatively display many of the features observed in the coupled Rijke tube system. This can be explained by the similarity in the slow flow amplitude equations of the two systems. 

Thus, we have analytically and numerically determined that varying system and coupling parameters shifts the Hopf points of the oscillators without altering their criticality. As a result, we observed explosive hysteretic transitions between LCO and AD states on delay coupling two Rijke tube oscillators which individually exhibit subcritical Hopf bifurcation. Apart from AD, we also analytically explained the occurrence of phase-flip bifurcation and its associated frequency trends in the system. Having examined the dynamics of the delay coupled system when both the Rijke tubes have the same length ($l$) and thus the same system parameters (i.e., the same natural frequency and amplitude in the uncoupled state), we next introduce mismatch in the length of the Rijke tubes ($l_A\neq l_B$) into the model and investigate its influence on the synchronization and amplitude suppression behavior of the system. 

\subsection{\label{sec:IIIB}Analysis of non-identical delay coupled Rijke tube oscillators}

From Fig.~\ref{fig:4}(a) in Sec.~\ref{sec:IIIA}, we noticed that for low values of coupling strength, mutual delay coupling is insufficient to completely suppress high amplitude limit cycle oscillations (say, $\overline{W} = 0.59$) of the acoustic field in a pair of identical Rijke tube oscillators. Recently, Dange \textit{et al.} \cite{dange2019oscillation} and Premraj \textit{et al.} \cite{premraj2021effect} demonstrated that such oscillations can be quenched by introducing a mismatch in the system parameters, such as the natural frequencies and amplitudes of the oscillators in the uncoupled state. We now introduce a small mismatch in the length of the Rijke tube oscillators. We keep the length of oscillator A ($l_A=l$) constant, whereas the length of oscillator B ($l_B$) is varied (see Fig.~\ref{fig:1}). Thus, we now consider the case when $r \neq 1$ in the governing equations of the delay coupled Rijke tubes [Eqs.~\eqref{eq:gov1}-\eqref{eq:gov4}]. 

We use the measure $\alpha=r-1=\left(l_B-l_A\right)/l_A$, henceforth referred to as `mismatch parameter', to quantify the mismatch in the lengths of the Rijke tubes. A positive $\alpha$ indicates that Rijke tube B is lengthened with respect to Rijke tube A, while a negative $\alpha$ implies shortening of Rijke tube B. By changing the length of an isolated Rijke tube, we bring about a change in the dimensional natural frequency and also the amplitude of the LCOs in it. Increasing the length of a Rijke tube decreases its dimensional natural frequency and the non-dimensional heater location (since the dimensional heater location is constant in the Rijke tube); as a result, this leads to a decrease in the amplitude of the limit cycle oscillations \cite{subramanian2010bifurcation, gopalakrishnan2014influence}. Thus, in the model of delay coupled non-identical Rijke tubes, the longer Rijke tube possesses LCOs of smaller amplitude in its uncoupled state. Next, we investigate how the introduction of mismatch parameter ($\alpha$) affects the amplitude of LCOs in the system. 

\subsubsection{\label{sec:IIIB1}Amplitude death and partial amplitude death in delay coupled non-identical Rijke tube oscillators}

\begin{figure*}
\includegraphics[width=17.0 cm]{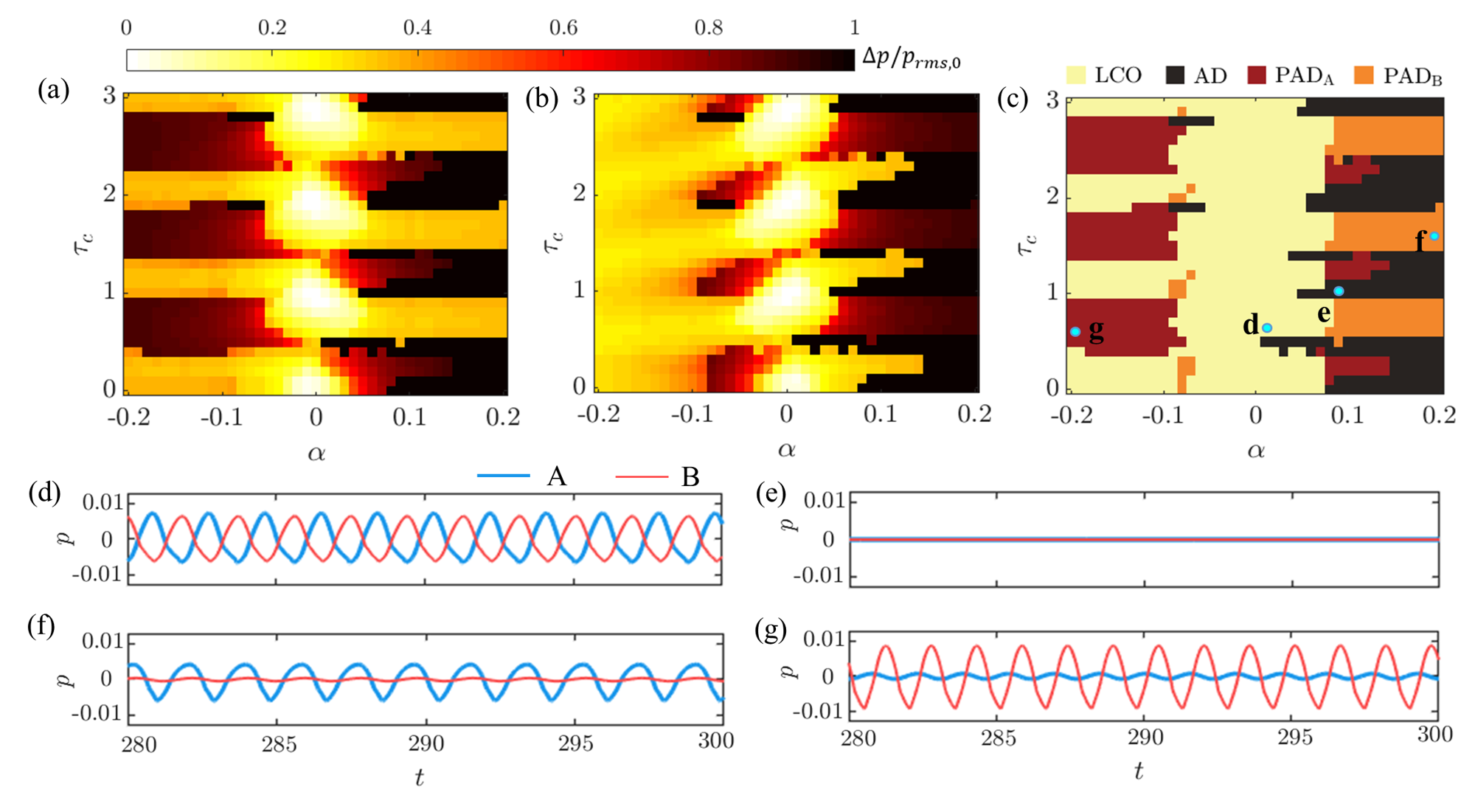}
\caption{\label{fig:7} Color maps showing the effect of variation in mutual coupling delay ($\tau_c$) and mismatch parameter ($\alpha$) on the relative amplitude suppression ($\Delta p/p_{rms,0}$) of (a) oscillator A and (b) oscillator B of the system of delay coupled non-identical Rijke tube oscillators. (c) Two-parameter bifurcation plot between $\tau_c$ and $\alpha$ illustrates the various states of coupled dynamics in the system. The coupled behavior of the system is asymmetric about the mismatch parameter. The temporal variations of the non-dimensional acoustic pressure oscillations for both the oscillators A and B are illustrated for the case of (d) LCO, (e) AD, (f) PAD\textsubscript{B}, and (g) PAD\textsubscript{A}, with the corresponding points marked in (c). $\overline{W}=0.59$ and $K_{\tau c}=0.15$ are fixed for all plots.}
\end{figure*}

In Figs.~\ref{fig:7}(a) and \ref{fig:7}(b), we examine the effect of varying the coupling delay ($\tau_c$) and the mismatch parameter ($\alpha$) on the amplitude suppression behavior of oscillator A and oscillator B, respectively, for constant values of normalized heater power ($\overline{W}$) and coupling strength ($K_{\tau c}$). Complete suppression ($\Delta p/p_{rms,0}=1$) and a lack thereof ($\Delta p/p_{rms,0}=0$) are indicated by dark and light zones, respectively. In the absence of mismatch ($\alpha= 0$), we note that the oscillations in both the oscillators are not quenched for the given values of heater power and coupling strength. However, the addition of finite mismatch results in better suppression of LCOs in one or both the oscillators. Following the work by Dange \textit{et al.} \cite{dange2019oscillation}, depending on whether LCOs in either A, B or both the oscillators are quenched, we classify the coupled behavior of Rijke tube oscillators into four distinct dynamical states, which are (i) limit cycle oscillations (LCO), (ii) amplitude death (AD), (iii) partial amplitude death in oscillator B (PAD\textsubscript{B}), and (iv) partial amplitude death in oscillator A (PAD\textsubscript{A}). The two-parameter bifurcation diagram between $\alpha$ and $\tau_c$ in Fig. ~\ref{fig:7}(c) illustrates the occurrence of these four states in the delay coupled system. The temporal variations of the acoustic pressure corresponding to these states [marked by points d, e, f, and g in Fig.~\ref{fig:7}(c)] are illustrated in Figs.~\ref{fig:7}(d) to \ref{fig:7}(g).

\begin{figure*}[t]
\includegraphics[width=14cm]{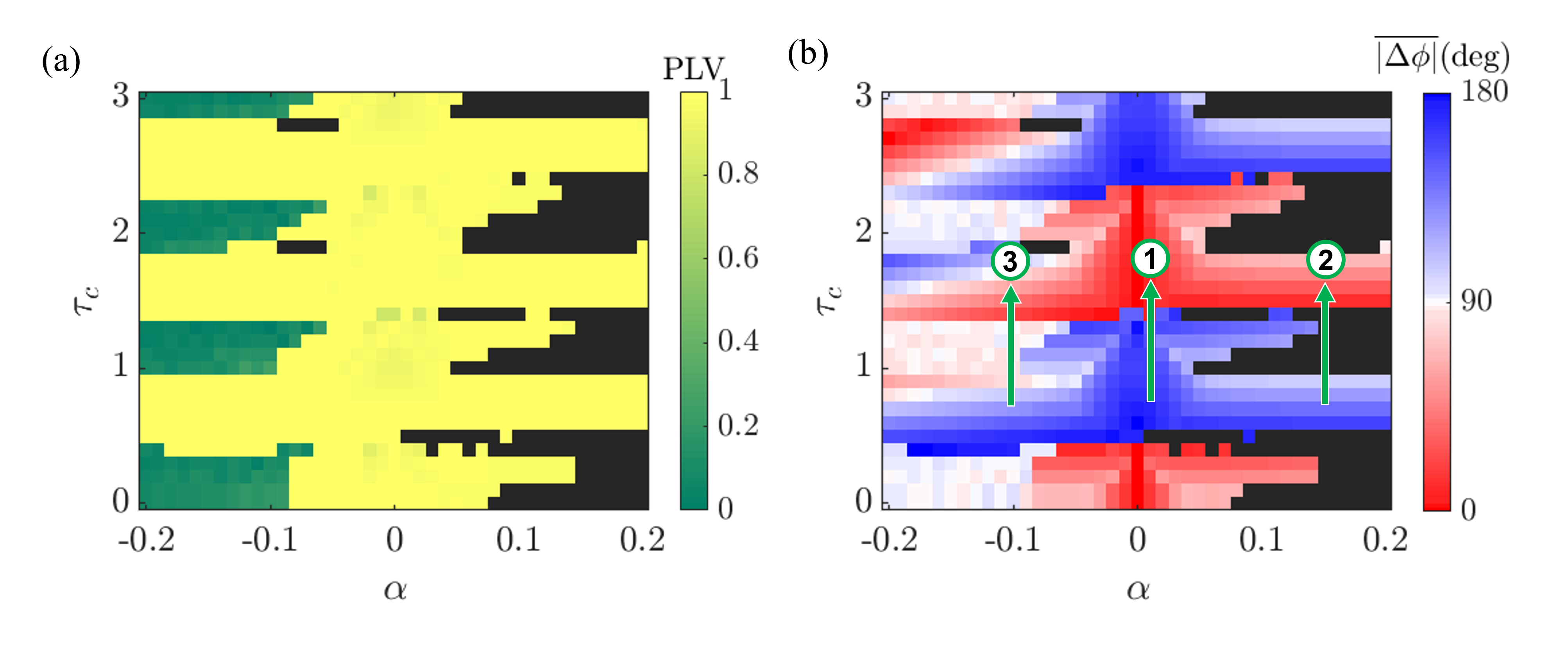}
\caption{\label{fig:8} Two-parameter bifurcation diagrams between the coupling delay ($\tau_c$) and the mismatch parameter ($\alpha$) where the color maps depict variation in (a) the phase-locking value (PLV) and the (b) the mean phase difference ($\overline{|\Delta\phi|}$) between the oscillations in the system of delay coupled non-identical Rijke tube oscillators. In (b), the arrows indicate the three routes through which non-identical delay coupled Rijke tube oscillators transition between in-phase and anti-phase synchronized state. These are (1) via phase-flip bifurcation, (2) via an intermediate state of AD, and (3) via an intermediate state of desynchronized LCOs. $\overline{W}=0.59$ and $K_{\tau c}=0.15$ are fixed for both the plots. The black regions denote AD state, where PLV and $\overline{|\Delta\phi|}$ are not defined.}
\end{figure*}

As previously mentioned in Sec.~\ref{sec:IIIA}, the system is said to have achieved AD state [depicted in Fig.~\ref{fig:7}(e)] if the oscillations in both the Rijke tubes are quenched after coupling the oscillators. The dynamical state wherein one of the oscillators exhibits a nearly quenched state (described by small-amplitude oscillations) and the other oscillator of the coupled system exhibits large amplitude oscillations is referred to as partial amplitude death (PAD) \cite{atay2003total,koseska2013oscillation,dange2019oscillation}. In Sec.~\ref{sec:IIIA}, we asserted how a Rijke tube cannot maintain its steady state when it is coupled to another Rijke tube exhibiting LCOs due to coupling-induced periodic oscillations [Figs.~\ref{fig:2}(b) and \ref{fig:2}(c)]. Hence, during the state of partial amplitude death (PAD), the oscillations in one of the Rijke tubes are greatly suppressed and their amplitude is small as compared to the LCOs in the other Rijke tube. From Figs.~\ref{fig:7}(a) and \ref{fig:7}(b), we observe that for large negative values of $\alpha$, the relative amplitude suppression ($\Delta p/p_{rms,0}$) in oscillator A rises to about 80\% while oscillator B still exhibits high amplitude LCOs. We refer to this state as partial amplitude death in oscillator A (PAD\textsubscript{A}), illustrated in Fig.~\ref{fig:7}(g). Similarly, we say that the system is in a state of partial amplitude death in oscillator B (PAD\textsubscript{B}), depicted in Fig.~\ref{fig:7}(f), when oscillations in Rijke tube B are quenched by at least 80\% (i.e., $\Delta p/p_{rms,0}\geq80\%$) while Rijke tube A exhibits high amplitude LCOs. When the oscillations in either of the oscillators are not significantly quenched ($\Delta p/p_{rms,0}<80\%$ in both the oscillators), we assign the state as LCO (i.e., limit cycle oscillations). All amplitude measurements are acquired after a sufficiently long time such that the transients are negligible. 

From the two-parameter bifurcation plot [Fig.~\ref{fig:7}(c)], we observe LCOs in both the oscillators for lower magnitudes of mismatch. We note that, on further increasing the magnitude of $\alpha$, the system attains AD or PAD state for recurring ranges of coupling delay ($\tau_c$). Varying $\alpha$ strongly affects the amplitude of oscillations in Rijke tube oscillator B (whose length is varied). For high values of $\alpha$, the oscillations in oscillator B are substantially suppressed regardless of the value of $\tau_c$. On the other hand, the suppression of oscillations in oscillator A (whose length is kept constant) is more affected by changes in coupling delay than by variation in $\alpha$. In general, we notice that the oscillations in the longer tube (which has smaller amplitude of LCOs in the uncoupled state) are quenched better. As a result, we observe large regions of PAD\textsubscript{B} on the positive side and PAD\textsubscript{A} on the negative side of $\alpha$ in the bifurcation diagram shown in Fig.~\ref{fig:7}(c). We see significantly larger regions of AD for positive mismatch as compared to negative mismatch. This matches well with our analytical approximation discussed towards the end of this section, where we predict that lengthening oscillator B, i.e., setting $\alpha$ to a positive value, promotes the occurrence of AD in the delay coupled system. The occurrence of PAD\textsubscript{A} and PAD\textsubscript{B} in two coupled non-identical Rijke tubes was experimentally demonstrated by Dange \textit{et al.} \cite{dange2019oscillation} and Sahay \textit{et al.} \cite{sahay2021dynamics}.

\subsubsection{\label{sec:IIIB2} Different routes between synchronization states and to AD in delay coupled non-identical Rijke tube oscillators}

Having discussed the trends in amplitude suppression, we will now examine the routes through which the system of delay coupled non-identical Rijke tube oscillators transition between in-phase and anti-phase synchronized states. Towards this purpose, we track the phase locking value (PLV) and the mean phase difference ($\overline{|\mathrm{\Delta\phi}|}$) between the oscillations of the Rijke tubes during LCO and PAD states as per Fig.~\ref{fig:7}(c). 

Phase-locking value (PLV) measures the level of synchronization between the two Rijke tube oscillators and is given by the following expression \cite{pikovsky2003synchronization}:
\begin{align}
\text{PLV} = \frac{1}{n} \left\lvert \sum_{j=1}^n \exp(i\Delta\phi) \right\rvert,
\label{eq:plv}
\end{align}
where $n$ is the length of the acoustic pressure signal and $\Delta\phi$ is the instantaneous phase difference between the acoustic pressure signals in the two Rijke tubes.
The value of PLV ranges from 0 to 1, with zero indicating desynchronization and 1 indicating synchronization of the oscillators. We do not calculate PLV for the AD state due to the absence of oscillations in the oscillators, in which case PLV does not have any physical meaning. 
\begin{figure*}[t]
\includegraphics[width=16cm]{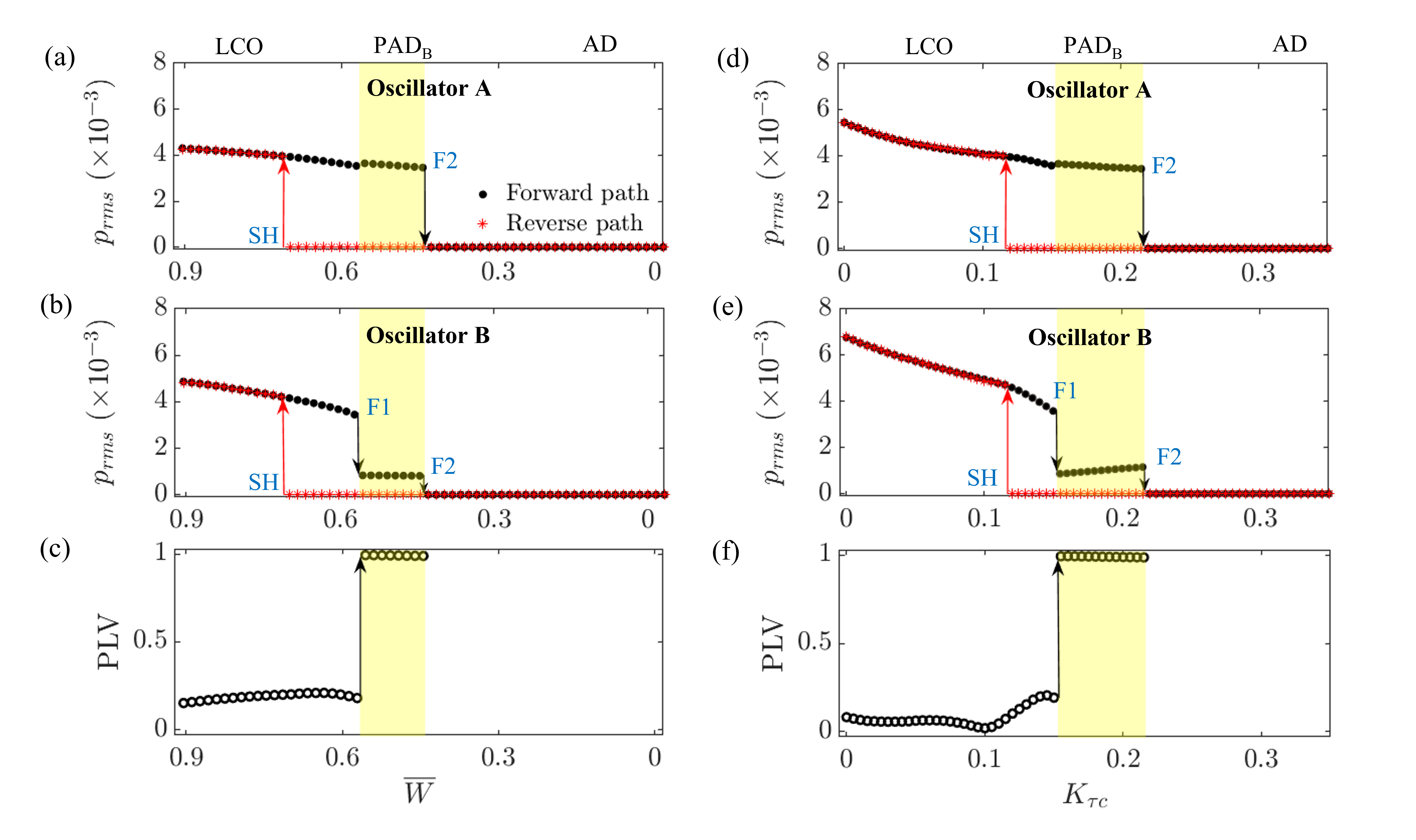}
\caption{\label{fig:9} Variation of root-mean-square value of acoustic pressure oscillations, $p_{rms}$, with normalized heater power, $\overline{W}$, during the transition from desynchronized LCO to AD, both in the forward (decreasing $\overline{W}$) and reverse (increasing $\overline{W}$) paths for (a) oscillator A and (b) oscillator B in a system of two delay coupled non-identical Rijke tube oscillators. (c) The corresponding variation in the phase-locking value (PLV) with $\overline{W}$ for the forward path (decreasing $\overline{W}$) shows sudden synchronization of the limit cycle oscillations during the state of partial amplitude death (PAD\textsubscript{B}), before the system attains amplitude death (AD). Similar variations of $p_{rms}$ and PLV as a function of the coupling strength, $K_{\tau c}$, are presented in (d)-(f). `SH' indicates subcritical Hopf bifurcation, while `F1' and `F2' indicate fold bifurcations. Region of PAD state is highlighted in yellow. $K_{\tau c} = 0.15$ is fixed for plots (a)-(c), while $\overline{W} = 0.59$ is fixed for plots (d)-(f). $\alpha = -0.1$ and $\tau_c = 0.2$ are fixed in all plots.}
\end{figure*}

From the bifurcation diagram [Fig.~\ref{fig:8}(a)], we observe that for most values of the mismatch parameter ($\alpha$), the oscillations are synchronized. However, for around $\alpha < -0.08$, the oscillations are desynchronized when they are in the LCO state. Figure~\ref{fig:8}(b) shows that the synchronized regions further comprises alternate bands of in-phase (IP) synchronization and anti-phase (AP) synchronization of oscillators A and B on variation in the $\tau_c$. Based on the value of the mismatch parameter, we observe three distinct routes through which transitions between IP and AP states occur in the system on variation of the coupling delay ($\tau_c$). Firstly, for small magnitudes of mismatch [marked by the arrow `1' in Fig.~\ref{fig:8}(b)], the transitions between IP and AP states are sudden, indicating PFB. The second way of transitioning between IP and AP states is through an intermediate state of AD. This route is observed for comparatively larger positive values of mismatch parameter [depicted by the arrow `2' in Fig.~\ref{fig:8}(b)]. We examined these two routes previously in identical delay coupled oscillators in Sec.~\ref{sec:IIIA} [refer Figs.~\ref{fig:3}(a) and \ref{fig:4}(a)]. Here, from Figs.~\ref{fig:8}(a) and \ref{fig:8}(b), we see that non-identical Rijke tube oscillators can also transition between IP and AP states through a third route, which is via an intermediate state of desynchrony. This route is mainly observed for large negative values of mismatch parameter [around $\alpha < -0.08$, depicted by the arrow `3' in Fig.~\ref{fig:8}(b)]. We find the transition between synchronized and desynchronized states on variation of $\tau_c$ to be abrupt, as indicated by the discontinuous change in color in Fig.~\ref{fig:8}(a).

Next, we examine the route to AD on variation of the normalized heater power, $\overline{W}$, for delay coupled non-identical Rijke tube oscillators exhibiting desynchronized LCOs. The one-parameter bifurcation diagrams in Figs.~\ref{fig:9}(a) and \ref{fig:9}(b) depict how varying $\overline{W}$ affects the root-mean-square value of acoustic pressure oscillations ($p_{rms}$) in coupled oscillators A and B, respectively. The values of $K_{\tau c}$, $\tau_c$ and $\alpha$ are chosen as 0.15, 0.2 and -0.1, respectively, so that we get desynchronized limit cycle oscillations at high values of $\overline{W}$ according to Fig.~\ref{fig:8}(a). We observe that, on decreasing $\overline{W}$ in the forward path, the desynchronized oscillations undergo secondary fold bifurcation [denoted as `F1' in Fig.~\ref{fig:9}(b)], where the amplitude of oscillator B slightly increases whereas that of oscillator A drops to a very low value. We refer to this state as partial amplitude death. We also note from Fig~\ref{fig:9}(c) that this sudden change in the amplitude of the acoustic oscillations in the forward path is accompanied by a jump in the PLV between the oscillators to one, indicating synchronization of the oscillators. Decreasing $\overline{W}$ further causes both the oscillators to attain AD through another fold bifurcation [denoted as `F2' in Figs.~\ref{fig:9}(a) and \ref{fig:9}(b)]. Thus, on lowering the heater power (a system parameter), desynchronized LCOs in delay coupled non-identical Rijke tube oscillators are quenched to AD state through an intermediate state of PAD wherein the oscillations are synchronized. In the reverse path, the transition from AD to desynchronized LCO state occurs directly through a subcritical Hopf bifurcation [marked as `SH' in Figs~\ref{fig:9}(a) and~\ref{fig:9}(b)]. 

Figures~\ref{fig:9}(d)-\ref{fig:9}(f) show similar behavior on variation of the coupling strength, $K_{\tau c}$. Increasing the value of coupling strength, $K_{\tau c}$ first causes desynchronized oscillations to synchronize during the state of PAD. Further increase in $K_{\tau c}$ leads to AD state. On the other hand, the desynchonized oscillations are restored in the system in the reverse path without an intermediate state of PAD. Interestingly, synchronized limit cycle oscillations are quenched on decreasing $\overline{W}$ or increasing $K_{\tau c}$ without an intermediate PAD state, which is similar to the results of identical oscillators discussed in Figs.~\ref{fig:3} and \ref{fig:4} in Sec.~\ref{sec:IIIA}. Furthermore, in Fig.~\ref{fig:9}, we observe hysteresis between amplitude death and oscillatory states for delay coupled non-identical Rijke tube oscillators.

\subsubsection{\label{sec:IIIB3} Analytical approximation for delay coupled non-identical Rijke tube oscillators}

We will now analytically examine the effect of the mismatch parameter on the occurrence of AD in the system of delay coupled non-identical Rijke tube oscillators. Towards this, we follow a methodology similar to what we utilized in Sec.~\ref{sec:IIIA3} (detailed in Sec.~III of the Supplementary Material) for delay coupled identical oscillators. Accordingly, we linearize Eqs.~\eqref{eq:gov1}-\eqref{eq:gov4} while considering only the first mode and subsequently use the method of averaging with an additional assumption of infinitesimal value of $\alpha$. Through this analysis, we find the condition for achieving AD in the system of delay coupled non-identical oscillators to be the following:
\begin{align}
\left|\cos\left(\omega\tau_c\right)\right|<& \Bigg[1 - \frac{(\sigma/\omega) \sin\left(\omega\tau_h\right)-b_0}{K_{\tau c} \sin^2(k x_c)} \nonumber\\
&+\alpha\left(\frac{(\beta+\epsilon)\sigma/\omega \sin\left(\omega\tau_h\right)-\epsilon b_0}{2K_{\tau c}\sin^2(k x_c)} \right) \Bigg],
\label{eq:20}
\end{align}
where $\beta=1+2kx_f\cot(2kx_f)$, $\epsilon = 3/2 - k x_c \cot(k x_c)$ and other parameters are as given in Eq.~\eqref{eq:8}. In Eq.~\eqref{eq:20}, we now focus only on the mismatch parameter, $\alpha$. We see that increasing the value of $\alpha$ in the positive direction, i.e., lengthening oscillator A, widens the range of parameters over which AD can be achieved [since the right-hand-side of Eq.~\eqref{eq:20} increases]. On the other hand, a negative value of $\alpha$, i.e., decreasing the length of oscillator A, seems to decrease the range of parameters for attaining AD. In Sec.~\ref{sec:IIIA3}, we found that Eq.~\eqref{eq:12} is indicative of the Hopf point of delay coupled identical oscillators. Similarly, here Eq.~\eqref{eq:20} determines the Hopf point of the two delay coupled non-identical oscillators. Note that, despite the presence of parameter mismatch, the oscillators share the same Hopf point when they are delay coupled. 

Thus, we analytically infer that the addition of frequency detuning and amplitude mismatch, achieved by lengthening one Rijke tube oscillator while keeping the length of the other oscillator constant, in a delay coupled system can result in the occurrence of amplitude death. This inference matches with our numerical results in Fig.~\ref{fig:7}(c) and previous experimental results by Dange \textit{et al.} \cite{dange2019oscillation}. In Fig.~\ref{fig:7}(c), we observe a few small islands of AD for negative values of mismatch. However, due to the assumption of small coupling delay, our analysis does not capture this trend. A more rigorous analysis which does not make the simplifying assumptions of small amplitude and small magnitudes of mismatch parameter is required to predict the stability of limit cycles and to completely explain the presence of PAD and desynchronization, which presents a scope for future study. 

\section{\label{sec:IV}Conclusions}

In this study, we investigated the occurrence of synchronization and amplitude suppression in a model of two coupled Rijke tube oscillators. We shed light on how system parameters (such as the amplitude and the frequency of the oscillations in the uncoupled state) and coupling parameters (such as the coupling strength and the coupling delay) affect the dynamical behavior of the system. Through approximate analytical solutions and numerical simulations, we demonstrated the occurrence of synchronization and amplitude death (AD) in two delay coupled identical Rijke tube oscillators. We observed that the nature of transition to AD for coupled Rijke tube oscillators is dependent on the criticality of the bifurcation of the individual oscillators; the transition is explosive (first-order) for oscillators that individually exhibit subcritical Hopf bifurcation, while it is continuous (second-order) when the individual oscillators exhibit supercritical Hopf bifurcation. We also observed two states of synchronized oscillations, i.e., in-phase and anti-phase synchronization, and the transition between these states happens either via an intermediate state of AD or through phase-flip bifurcation (PFB) on increasing the value of coupling delay. We analytically predicted the critical values of coupling and system parameters for achieving AD and PFB in Rijke tube oscillators that are delay coupled. As compared to an isolated Rijke tube oscillator, we observed that delay coupling shifts forward the Hopf points of the Rijke tube oscillators without altering their criticality. Furthermore, we showed that oscillations can be induced in a damped oscillator by coupling it with another oscillator exhibiting limit cycle oscillations (LCOs). 

The introduction of the mismatch parameter, i.e., a small mismatch in the length of the Rijke tubes, causes a mismatch in the natural frequencies and the amplitudes of the oscillators. We observed the introduction of mismatch parameter to suppress high amplitude LCOs, resulting in the occurrence of multiple parametric regions of partial amplitude death (PAD) and AD in the system of delay coupled Rijke tube oscillators. We discovered desynchronized oscillations as an intermediate state between IP and AP for large negative values of the mismatch parameter. We further found the transition from desynchronized oscillations to AD to happen through an intermediate state of PAD on varying both system and coupling parameters. The presence of synchronization, PFB, AD, and PAD in our model corroborates the experimental observations by Dange \textit{et al.} \cite{dange2019oscillation}. 

We thus examined in detail a model that captures all of the dynamical phenomena observed experimentally in coupled thermoacoustic oscillators. We also demonstrated the important role played by system parameters in determining the dynamical state of coupled limit cycle oscillators, and detailed the mechanisms through which these dynamical changes can occur. The findings in this study may provide insights into the coupled behavior of acoustic fields of practical combustion systems, such as can, can-annular or annular combustors. This could in turn help us devise control strategies to mitigate thermoacoustic instability in these systems, which is a rarely explored area of research till date. Examples of such control strategies may include using connecting tubes of appropriate lengths and diameters to acoustically connect cans of the engine or introducing mismatch in the length of the coupled combustion chambers. The effectiveness of these control strategies first needs to be rigorously tested experimentally and theoretically in coupled turbulent systems before their implementation in real engines in the future.

\begin{acknowledgments}
S. S. is thankful to the support offered by Prof. Preeti Aghalayam and other members of the Young Research Fellow Program of Indian Institute of Technology Madras (Project ID: 202025), India. R. I. S. gratefully acknowledges the IoE initiative (SB/2021/0845/AE/ MHRD/002696), and the J. C. Bose Fellowship (No. JCB/2018/000034/SSC) from the Department of Science and Technology (DST) for the financial support. The authors are grateful to Mr. A. Sahay for several fruitful discussions.
\end{acknowledgments}

\bibliography{Manuscript}

\end{document}


\preprint{AIP/123-QED}

\title{Supplementary Material\\
Dynamical States and Bifurcations in Coupled Thermoacoustic Oscillators}

\author{Sneha Srikanth}
 \affiliation{%
 Department of Mechanical Engineering, Indian Institute of Technology Madras, Chennai 600036, India
 }%
\author{Samadhan A. Pawar}%
\affiliation{%
 Department of Aerospace Engineering, Indian Institute of Technology Madras, Chennai 600036, India
}%
 \email{samadhanpawar@ymail.com}
\author{Krishna Manoj}%
\affiliation{%
Department of Mechanical Engineering, Massachusetts Institute of Technology, Cambridge, MA 02139, USA
}%
\author{R. I. Sujith}%
\affiliation{%
 Department of Aerospace Engineering, Indian Institute of Technology Madras, Chennai 600036, India
}%
\date{\today}

\maketitle
\renewcommand{\theequation}{S\arabic{equation}}
\renewcommand{\thefigure}{S\arabic{figure}} 

In this supplementary material, we present additional results that supplement our findings in the main manuscript. We validate the generality of our findings by illustrating the shift in the Hopf point due to delay coupling in oscillators exhibiting supercritical Hopf bifurcation. We also compare some of the numerical results with previous experimental observations by Dange \textit{et al.} \protect\cite{dange2019oscillation}. Finally, we present a detailed derivation of the analytical approximation of the delay coupled Rijke tube oscillators. 

\section{\label{appa}Effect of delay coupling on oscillators exhibiting supercritical Hopf bifurcation}

In Fig.~2 of Sec.~III A of the main manuscript, we studied how delay coupling shifts the subcritical Hopf points of Rijke tube oscillators. Subramanian \textit{et al.} \protect\cite{subramanian2010bifurcation} explained how the present model of the Rijke tube can only exhibit subcritical and not supercritical Hopf bifurcation due to the square-root nonlinearity of the source term. However, experiments by Etikyala and Sujith \protect\cite{etikyala2017change} show that for low flow rates, the Rijke tube undergoes supercritical Hopf bifurcation. Therefore, in this section, we consider the model of two identical thermoacoustic oscillators exhibiting supercritical Hopf bifurcation and examine how delay coupling affects their Hopf points. The governing equations Eqs.~(5)-(9) and (11) of the main manuscript remain the same, apart from maintaining $r=1$, while Eqs.~(10) and (12) are modified as below:
\begin{align}
\label{eq:A1}
    {{\dot{P}}_j}^{A,B} + 2\zeta_j\omega_j{P_j}^{A,B} &- k_j{U_j}^{A,B}  \nonumber \\ &= \underbrace{\frac{\sqrt{3}}{2} W \sin(kx_f)\Big( u_f^{A,B}(t-\tau_h) - \frac{3}{4} u_f^{A,B}(t-\tau_h)^2 - \frac{9}{8} u_f^{A,B}(t-\tau_h)^3 \Big)}_{\text{Source}} \nonumber\\ &\quad  +  \underbrace{\dfrac{K_{\tau c}}{\gamma M}\sin(kx_c) \Big(p_c^{B,A}(t-\tau_c)-p_c^{A,B}(t)\Big)}_{\text{Delay coupling}}
\end{align}
We obtained the source term of Eq.~(\ref{eq:A1}) by expanding the nonlinearity in Eqs.~(10) and (12) of the main manuscript upto the cubic term and then changing the sign of the cubic term so that the Hopf bifurcation is supercritical in nature \protect\cite{subramanian2010bifurcation}. For our subsequent analysis, we fix some of the model parameters according to Table~1 in Sec.~II. 

\begin{figure}
\includegraphics[width=14cm]{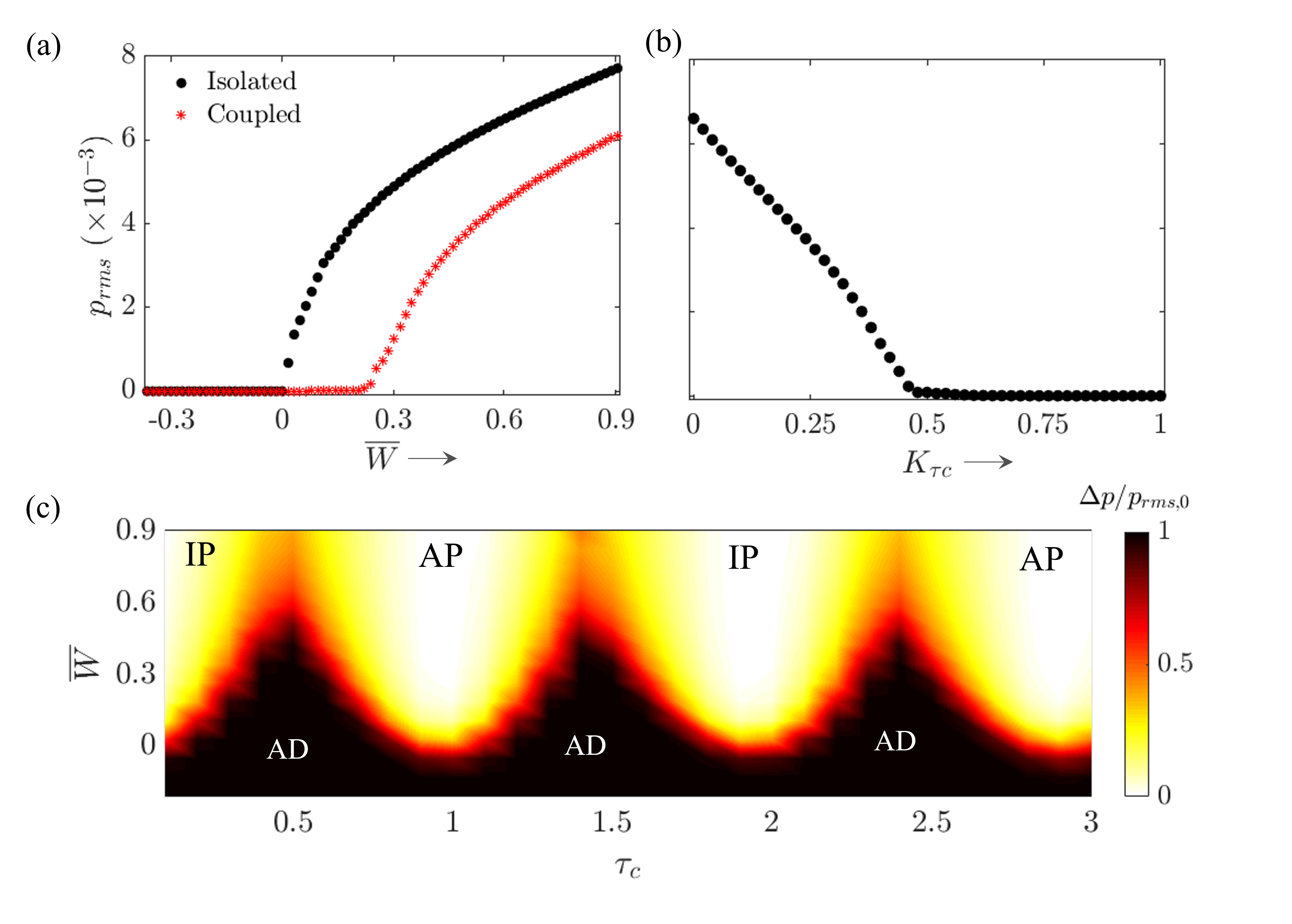}
\caption{\label{fig:A1}(a) One-parameter bifurcation plot between the root-mean-square value of the acoustic pressure oscillations, $p_{rms}$, and the normalized heater power, $\overline{W}$, shows that delay coupling shifts the supercritical Hopf point of thermoacoustic oscillators to a higher value of $\overline{W}$ for $\tau_c = 1.3$ as compared to when they are isolated. (b) One-parameter bifurcation plot between $p_{rms}$ and coupling strength, $K_{\tau  c}$, depicts continuous (second-order) transition of limit cycle oscillations to amplitude death (AD) for $\overline{W}=0.59$ and $\tau_c = 0.7$. (c) Two-parameter bifurcation plot between $\overline{W}$ and coupling delay, $\tau_c$, illustrates the variation in the relative suppression of acoustic pressure oscillations ($\Delta p/p_{rms,0}$) and the existence of in-phase (IP) and anti-phase (AP) synchronization in two delay coupled thermoacoustic oscillators that individually exhibit supercritical Hopf bifurcation. $K_{\tau c}$ is fixed at 0.1 in plots (a) and (c)}
\end{figure}

From the one-parameter bifurcation plot in Fig.~\ref{fig:A1}(a), we observe that the thermoacoustic oscillator undergoes supercritical Hopf bifurcation at $\overline{W}=0$ in the absence of coupling. Here, $\overline{W}$ is the normalized heater power defined by $\overline{W} = W/W_H-1$, where $W_H=0.63$ is the heater power corresponding to the Hopf point of the oscillator. We find that delay coupling two thermoacoustic oscillators shifts forward their Hopf points (from $\overline{W} = 0$ to around $\overline{W} = 0.21$), similar to what we observe in Fig.~2(a) in Sec.~III A of the main manuscript. Furthermore, we observe that when the individual thermoacoustic oscillators of the coupled system exhibit supercritical Hopf bifurcation, the transition between the states of amplitude death (AD) and limit cycle oscillations (LCOs) is of second order and is not hysteretic. In Fig.~\ref{fig:A1}(b), we plot the variation of the root-mean-square value of the acoustic pressure fluctuations $p_{rms}$ during the transition from LCO to AD as the coupled strength ($K_{\tau c}$) is varied. We see that varying $K_{\tau c}$ results in continuous transition of LCOs to AD when the thermoacoustic oscillators have supercritical Hopf points in the uncoupled state.

In Fig.~\ref{fig:A1}(c), we present the two-parameter bifurcation diagram between the normalized heater power $\overline{W}$ and the coupling delay $\tau_c$ of the system. We observe the LCOs in the system to be suppressed to the greatest extent when $\tau_c$ is around  $1/2, 3/2, 5/2,$ …. We also observe alternate occurrence of in-phase (IP) and anti-phase (AP) synchronized regions in the delay coupled system on variation of $\tau_c$. The amplitude of the oscillations appears to dip when the system undergoes phase-flip bifurcation (PFB). All of these observations are similar to what we studied in Sec.~III~A [Figs.~4(a) and 5(c)] of the main manuscript.
\clearpage

\section{\label{appb}Comparison between model and experiments}
In this section, we compare some of the results obtained from the model of delay coupled Rijke tube oscillators in the current study with the corresponding experimental results obtained by Dange \textit{et al.}~\protect\cite{dange2019oscillation}. 

\begin{figure*}[b]
\includegraphics[width=13cm]{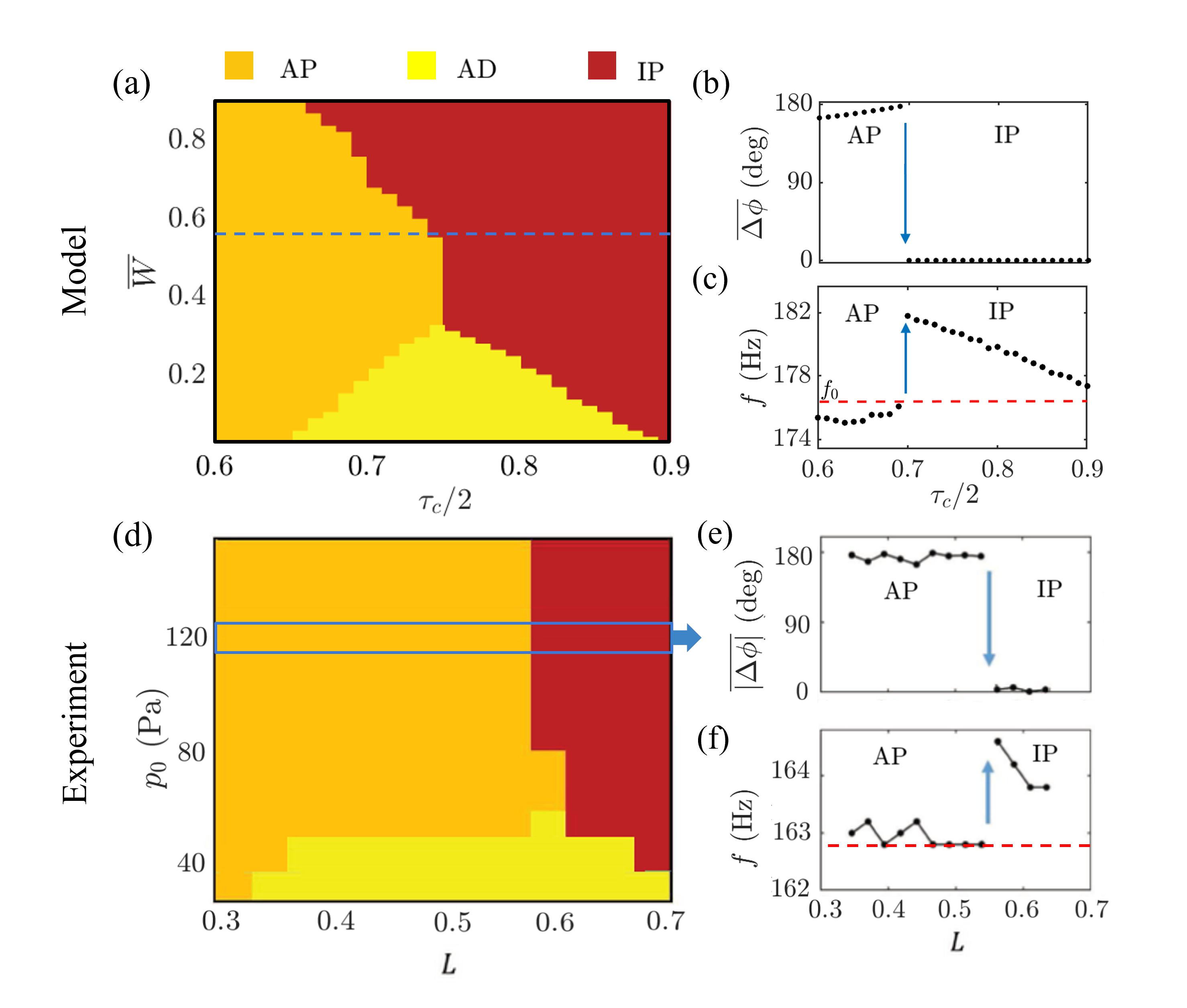}
\caption{\label{fig:B1}(a) A portion of the bifurcation diagram between the normalized heater power ($\overline{W}$) and mutual coupling delay ($\tau_c$) obtained from our model in Fig.~4(a) of the main manuscript. Trends in (b) the mean phase difference ($\overline{\left|\Delta\phi\right|}$) and (c) the dimensional dominant frequency ($f$) of the coupled oscillations on varying $\tau_c$. These plots are compared with the experimentally obtained (d) bifurcation diagram between the root-mean-square value ($p_0$) of the acoustic pressure oscillations in the uncoupled state and the non-dimensional length of connecting tube ($L$), and the variation of (e) $\overline{\left|\Delta\phi\right|}$ and (f) $f$ with $L$. In both (c) and (f), the red dashed line indicates the frequency of the oscillations in the uncoupled system. IP and AP indicate in-phase and anti-phase states of synchronization, respectively. Plots (d)-(f) are reproduced with permission from Dange \textit{et al.} \protect\cite{dange2019oscillation}.}
\end{figure*}

In Sec. III A of the main manuscript, we noted the occurrence of AD and PFB in the system of two mutually delay coupled identical Rijke tube oscillators [see Fig. 4(a) of the main manuscript]. We observed that any change in $\overline{W}$ directly corresponds to a change in the amplitude of the oscillations in the uncoupled state [refer Fig.~2(a)]. Hence, a portion of the bifurcation diagram [Fig.~4(a)] obtained from our model is presented in Fig.~\ref{fig:B1}(a) and is qualitatively compared with the experimentally obtained two-parameter bifurcation diagram [Fig.~\ref{fig:B1}(d)] between the root-mean-square value ($p_0$) of the acoustic pressure oscillations in uncoupled state and the length of the connecting tube ($L=l_c/\lambda$, where $l_c$ is the length of the connecting tube and $\lambda$ is the wavelength of the uncoupled oscillations) by Dange \textit{et al.}~\protect\cite{dange2019oscillation}. We also compare the corresponding trends in the dominant frequency ($f$) and the mean phase difference ($\overline{\left|\Delta\phi\right|}$) of the coupled system. From Fig.~\ref{fig:B1}, we observe a possible correlation between the non-dimensional quantities $\tau_c$ and $L$, which can be described as:
\begin{align}
\label{eq:B1}
  \tau_c & = \Tilde{\tau}_c/ (l/c_0), \nonumber\\
  \intertext{where $\Tilde{\tau}_c$ is the dimensional coupling delay. Since $\Tilde{\tau}_c$ is roughly $l_c/c_0$,} 
  \tau_c &\approx (l_c/c_0)/(l/c_0) \nonumber\\
&= l_c/l \approx l_c/(\lambda/2)=2L,    
\end{align}
where $l$ is the length of the Rijke tube and $c_0$ is the speed of sound. Hence, due to Eq.~(\ref{eq:B1}), we plot $\tau_c/2$ on the abscissa while presenting the results from our model. We observe a qualitative match between the results obtained from experiments and the model. In both the model and the experiments [refer to Figs.~\ref{fig:B1}(a) and \ref{fig:B1}(d), respectively], we observe the occurrence of AD for low amplitudes and phase-flip bifurcation (PFB) for high amplitudes LCOs in the uncoupled state. Additionally, in both the cases, the dominant frequency, $f$ [Figs.~\ref{fig:B1}(c) and \ref{fig:B1}(f)], and the phase difference, $\overline{\left|\Delta\phi\right|}$ [Figs.~\ref{fig:B1}(b) and \ref{fig:B1}(e)], between the oscillations in the two Rijke tube oscillators undergo an abrupt change during PFB. The dominant frequency falls after the occurrence of PFB on increasing the coupling delay in the model, or correspondingly, the length of connecting tube in the experiment. 

\begin{figure*}[t]
\includegraphics[width=14cm]{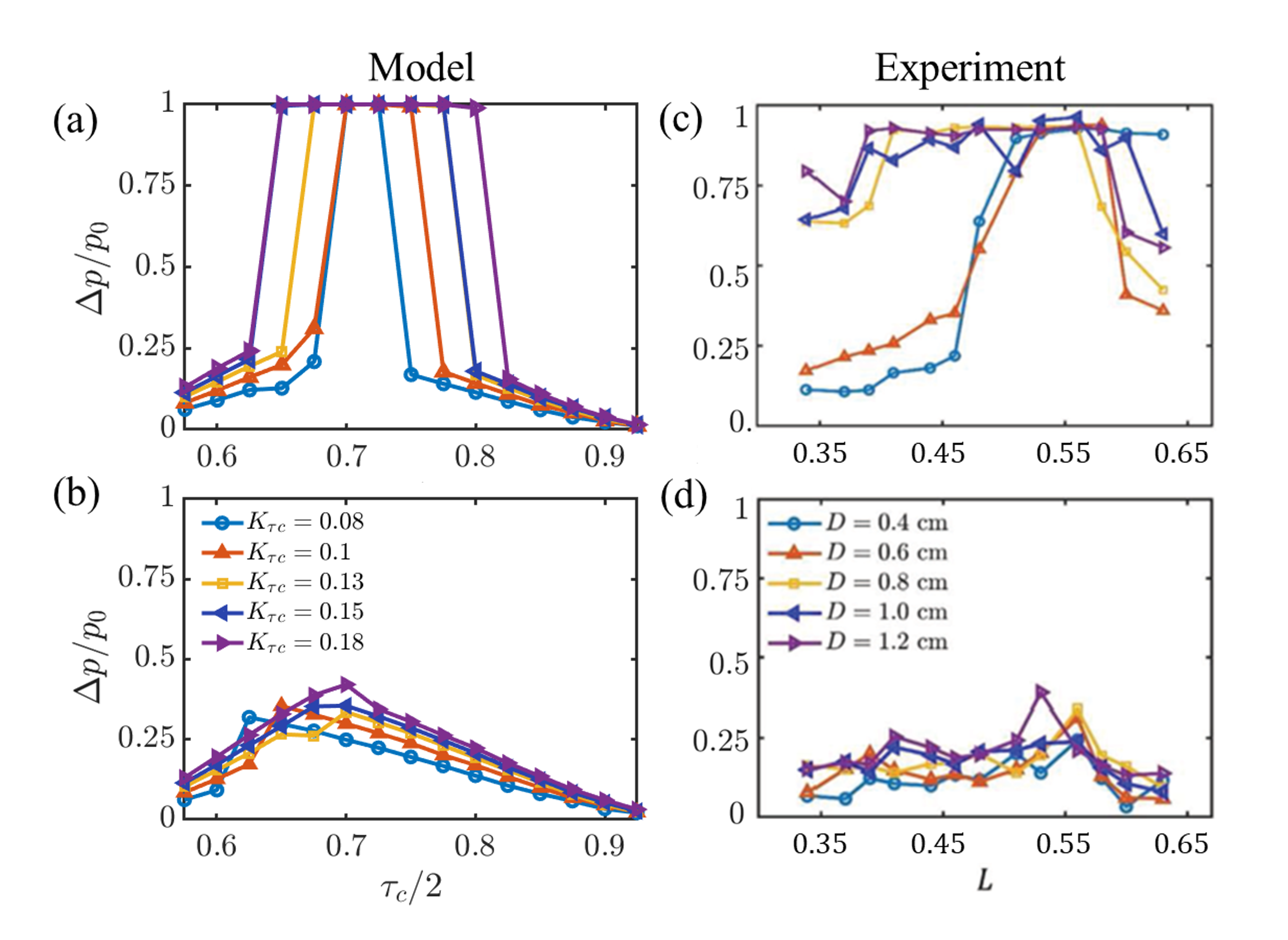}
\caption{\label{fig:B2} Comparison of the results obtained from our model with the corresponding experimental results shows possible correlation between the coupling parameters in the model and the experiments. Relative suppression of the amplitude of acoustic pressure oscillations ($\Delta p/p_0$) on varying $\tau_c$ (or $L$ in the experiment) for different values of $K_{\tau c}$ in the model (or different values of connecting tube diameter $D$ in the experiment). (a) $\overline{W}=0.12$ in the model (or $p_0=40$ Pa in the experiment) and (b) $\overline{W}=0.93$ (or $p_0=120$ Pa in the experiment). Plots (c) and (d) are reproduced with permission from Dange \textit{et al.} \protect\cite{dange2019oscillation}.}
\end{figure*}

Furthermore, we find that when the normalized heater power is low [Figs.~\ref{fig:B2}(a) and \ref{fig:B2}(c)], the anti-phase synchronized LCOs are completely suppressed (AD) for a particular range of $\tau_c$ or $L$, after which the system transitions to in-phase synchronization. The relative suppression of such oscillations in the model increases significantly on increasing the delay coupling strength ($K_{\tau c}$). A similar trend is observed in the experiment when the diameter of the connecting tube ($D$) is increased. 

For higher values of normalized heater power [Figs.~\ref{fig:B2}(b) and \ref{fig:B2}(d)], the given limited range of coupling strength (or diameter in the experiment) is not sufficient for the system to attain AD irrespective of the value of coupling delay (or correspondingly, the length of the connecting tube). The relative suppression of oscillations in the model in this case reaches a maximum at around $\tau_c/2= 0.7$, where the system undergoes PFB.

\newpage

\section{Analytical Approximation for Delay Coupled Identical Thermoacoustic Oscillators}
Here, we analytically predict the occurrence of amplitude death (AD) and phase-flip bifurcation (PFB) in two time-delay coupled identical thermoacoustic oscillators. We start with the governing equation Eq.~(15) of the main manuscript which considers only the first modes of the oscillators:
\begin{align}
\label{eq:S1}
    \ddot{U}_{A,B} + 2\zeta \omega \dot{U}_{A,B}+k^2 U_{A,B} + Wk \sin(kx_f)\Bigg[\sqrt{\Big|\frac{1}{3}+\cos(kx_f) U_{A,B} (t-\tau_h)\Big|}-\sqrt{\frac{1}{3}}\Bigg] \nonumber\\ + K_{\tau c} \sin^2(kx_c) [\dot{U}_{A,B}-\dot{U}_{B,A} (t-\tau_c)] = 0,
\end{align}
where the variables are defined as per the main manuscript. Expanding the nonlinear source term in the above equation as a Taylor series gives:
\begin{align}
\label{eq:S2}
\ddot{U}_{A,B}+b_0 \dot{U}_{A,B}+b_1 U_{A,B} + \sigma U_{A,B} (t-\tau_h) &+\sigma_2[U_{A,B} (t-\tau_h)]^2 +\sigma_3[U_{A,B} (t-\tau_h)]^3 \nonumber\\ &+ K_{\tau c} \sin^2(kx_c) [\dot{U}_{A,B}- \dot{U}_{B,A} (t-\tau_c)] =0,     
\end{align}
where $b_0 = 2\zeta\omega$, $b_1=k^2=\pi^2$, $\sigma=(\sqrt{3}/4)Wk \sin(2kx_f)$, $\sigma_2=-\frac{3}{4} \sigma \cos(k x_f)$ and $\sigma_3=\frac{9}{8} \sigma \cos^2 (k x_f)$. We have neglected higher order terms of the expansion in Eq.~(\ref{eq:S2}), with the assumption of small amplitudes. 

We will now perform the method of averaging on the above equation. Accordingly, we assume the solution to be of the form \protect\cite{balanov2008synchronization}:
\begin{eqnarray}
\label{eq:S3}
U_{A,B} = \frac{1}{2}[a(t)_{A,B}e^{i\omega t} + a^*_{A,B}(t)e^{-i\omega t}],
\end{eqnarray}
where $a_{A,B}(t)$ and $a^*_{A,B}(t)$ are complex conjugates of each other and are called the complex amplitudes of the oscillations. Since $a$ is a complex number and, thus, consists of two unknowns: the real and imaginary parts, while $U$ is a single unknown, we need a constraint equation. We use the following constraint equation for convenience \protect\cite{balanov2008synchronization}:
\begin{eqnarray}
\label{eq:S4}
\dot{a}_{A,B}e^{i\omega t} + \dot{a}_{A,B}^*e^{-i\omega t} = 0.
\end{eqnarray}
Differentiating Eq.~(\ref{eq:S3}) and applying the constraint equation Eq.~(\ref{eq:S4}) yields
\begin{eqnarray}
\label{eq:S5}
\dot{U}_{A,B} &=& \frac{i\omega}{2}(a_{A,B}e^{i\omega t} - a_{A,B}^*e^{-i\omega t} ), \\
\label{eq:S6}
\ddot{U}_{A,B} &=& i\omega\dot{a}_{A,B}e^{i\omega t}-\frac{\omega^2}{2}( a_{A,B}(t)e^{i\omega t} + a^*_{A,B}(t)e^{-i\omega t}).
\end{eqnarray}
Substituting Eqs.~(\ref{eq:S3}), (\ref{eq:S5}) and (\ref{eq:S6}) into Eq.~(\ref{eq:S2}),
\begin{align}
\label{eq:S7}
&i\omega\dot{a}_{A,B}e^{i\omega t} - \frac{\omega^2}{2}\left(a_{A,B}e^{i\omega t} + a^*_{A,B}e^{-i\omega t}\right) + \frac{i\omega(b_0 + K_{\tau c}\sin^2(kx_c))}{2} (a_{A,B}e^{i\omega t} - a^*_{A,B}e^{-i\omega t}) \nonumber\\ &+ \frac{b_1}{2}(a_{A,B}e^{i\omega t} + a^*_{A,B}e^{-i\omega t}) + \frac{\sigma}{2}\Big[a_{A,B}e^{i\omega (t-\tau_h)} + a^*_{A,B}e^{-i\omega(t-\tau_h)}\Big] + \frac{\sigma_2}{4}\Big[a^2_{A,B}e^{2i\omega (t-\tau_h)} \nonumber\\ &+ a^{*2}_{A,B}e^{-2i\omega (t-\tau_h)} + 2a_{A,B}a^*_{A,B}\Big] +\frac{\sigma_3}{8} \Big[ a^3_{A,B}e^{3i\omega (t-\tau_h)} + a^{*3}_{A,B}e^{-3i\omega (t-\tau_h)} + 3a^2_{A,B}a^*_{A,B}\nonumber\\ &\times e^{i\omega (t-\tau_h)}  + 3a_{A,B}a^{*2}_{A,B}e^{-i\omega (t-\tau_h)} \Big] - K_{\tau c} \sin^2(kx_c) \frac{i\omega}{2} \left(a_{B,A}e^{i\omega (t-\tau_c)} - a^*_{B,A}e^{-i\omega(t-\tau_c)}\right) = 0.
\end{align}
Here, we have assumed that $a_{A,B}(t)$ varies slowly with time, and the values of $\tau_h$ and $\tau_c$ are small; so, $a_{A,B}(t-\tau_h)$ and $a_{A,B}(t-\tau_c)$ are nearly equal to $a_{A,B}(t)$ \protect\cite{wahi2005study}. Dividing by $i\omega e^{i\omega t}$ and grouping terms,
\begin{align}
\label{eq:S8}
&\dot{a}_{A,B} + \Big(\frac{i\omega}{2} - \frac{ib_1}{2\omega}\Big) \left(a_{A,B} + a^*_{A,B}e^{-2i\omega t} \right) + \frac{b_0 + K_{\tau c}\sin^2(kx_c)}{2} \left(a_{A,B} - a^*_{A,B}e^{-2i\omega t} \right) - \frac{i\sigma}{2\omega}\Big[a_{A,B}
\nonumber\\ & e^{-i\omega \tau_h}  + a^*_{A,B}e^{-i\omega(2t - \tau_h)}\Big] - \frac{i\sigma_2}{4\omega} \Big[a^2_{A,B}e^{i\omega (t-2\tau_h)} + a^{*2}_{A,B}e^{-i\omega (3t-2\tau_h)} + 2a_{A,B}a^*_{A,B}e^{-i\omega t} \Big] - \frac{i\sigma_3}{8\omega} \nonumber\\ & \times \Big[ a^3_{A,B}e^{i\omega (2t-3\tau_h)} +   a^{*3}_{A,B}e^{-i\omega (4t-3\tau_h)} + 3a^2_{A,B}a^*_{A,B}e^{-i\omega\tau_h} +  3a_{A,B}a^{*2}_{A,B}e^{-i\omega (2t-\tau_h)} \Big] \nonumber\\ &- \frac{K_{\tau c}\sin^2(kx_c)}{2}\left(a_{B,A}e^{-i\omega\tau_c} - a^*_{B,A}e^{-i\omega(2t-\tau_c)}\right) = 0.
\end{align}

The aim of our calculations is to find the time evolution of the complex amplitude $a_{A,B}(t)$, which is a slow function of time as compared to the fast oscillations $e^{in\omega t}$, where $n$ is an integer. Hence, by averaging Eq.~(\ref{eq:S8}) over one time period of the fast oscillations (say, from $t=0$ to $t= 2\pi/\omega$), we can get rid of the fast oscillation terms since $\int_0^{2\pi/\omega}{e^{in\omega t}}=0$. Thus, we retain only the slow terms which remain nearly constant over the duration of a time period. Averaging Eq.~(\ref{eq:S8}), we get,
\begin{align}
\label{eq:S9}
\dot{a}_{A,B} + \Big[ \frac{i\omega}{2} - \frac{ib_1}{2\omega}\Big]a_{A,B} + \frac{b_0 + K_{\tau c}\sin^2(kx_c)}{2}a_{A,B} &- \frac{i\sigma}{2\omega}a_{A,B}e^{-i\omega \tau_h} -  \frac{3i\sigma_3}{8\omega}a^2_{A,B}a^*_{A,B}e^{-i\omega\tau_h} \nonumber\\ &-  \frac{K_{\tau c}\sin^2(kx_c)}{2}a_{B,A}e^{-i\omega\tau_c} = 0.
\end{align}

Let us derive the condition for attaining amplitude death (AD) for the system of delay coupled thermoacoustic oscillators. Towards this purpose, we cast Eq.~(\ref{eq:S9}) in matrix form, ignoring the nonlinear terms: 
\begin{gather}
 \begin{bmatrix} \dot{a}_A \\ \dot{a_B} \end{bmatrix}
 =
  \begin{bmatrix}
  -\frac{i\omega}{2} + \frac{ib_1}{2\omega} - \frac{b_0 + K_{\tau c}\sin^2(kx_c)}{2} + \frac{i\sigma}{2\omega}e^{-i\omega \tau_h} &
  \frac{K_{\tau c}\sin^2(kx_c)}{2}e^{-i\omega\tau_c} \\
  \frac{K_{\tau c}\sin^2(kx_c)}{2}e^{-i\omega\tau_c} &
  -\frac{i\omega}{2} + \frac{ib_1}{2\omega} - \frac{b_0 + K_{\tau c}\sin^2(kx_c)}{2} + \frac{i\sigma}{2\omega}e^{-i\omega \tau_h}
  \end{bmatrix}
  \begin{bmatrix} a_A \\ a_B \end{bmatrix} \\
  \intertext{or concisely written as,}
  \frac{d\textbf{a}}{dt} = \textbf{J}\textbf{a}.
\end{gather}

The system attains AD when the fixed point $a_A = a_B= 0$ gains stability. The fixed point is stable if the real parts of the eigenvalues of \textbf{J} are negative \protect\cite{strogatz2018nonlinear}. The eigenvalues of \textbf{J}, $\lambda_{1,2}$ are given by:
\begin{align}
    \lambda_{1,2} = \frac{1}{2}\Big[r \mp \sqrt{r^2 - 4\Delta}\Big],
\end{align}
where $r$ and $\Delta$ are the trace and determinant of \textbf{J}, respectively.
\begin{align}
   \lambda_{1,2} &= -\frac{i\omega}{2} + \frac{ib_1}{2\omega} - \frac{b_0 + K_{\tau c}\sin^2(kx_c)}{2} + \frac{i\sigma}{2\omega}e^{-i\omega \tau_h} \mp \frac{K_{\tau c}\sin^2(kx_c)}{2} e^{-i\omega\tau_c}& \nonumber\\
    &= - \frac{b_0 + K_{\tau c}\sin^2(kx_c)}{2} + \frac{\sigma}{2\omega} \sin{\omega \tau_h} \mp \frac{K_{\tau c}\sin^2(kx_c)}{2} \cos{\omega\tau_c} + \text{Imaginary terms}.&
\end{align}
Hence, to achieve AD, the real part of the largest eigenvalue must be less than zero, i.e.,
\begin{align}
\label{eq:S14}
\mp K_{\tau c}\sin^2(kx_c)\cos(\omega\tau_c) &< b_0 + K_{\tau c}\sin^2(kx_c) - \sigma/\omega \sin(\omega \tau_h)\nonumber\\
\Rightarrow |\cos(\omega\tau_c)| &< \frac{b_0 + K_{\tau c}\sin^2(kx_c) - \sigma/\omega \sin(\omega \tau_h)}{K_{\tau c}\sin^2(kx_c)}.
\end{align}
Equation \ref{eq:S14} gives the condition for achieving AD in the system. We will now resolve the complex amplitude into its real magnitude $R_{A,B}(t)$ and phase $\phi_{A,B}(t)$. Substituting $a_{A,B} = R_{A,B} e^{i\phi_{A,B}}$ into Eq.~(\ref{eq:S3}) gives us
\begin{eqnarray}
\label{eq:S15}
U_{A,B} &=& \frac{1}{2}[R_{A,B}e^{i\omega t + \phi_{A,B}} + R_{A,B}(t)e^{-i\omega t - \phi_{A,B}}] \nonumber\\
 &=& R_{A,B}\cos(\omega t + \phi_{A,B}).
\end{eqnarray}
Hence, $R_{A,B}(t)$ and $\phi_{A,B}(t)$ describe the amplitude and the phase of the oscillations in the oscillators of the coupled system. Substituting $a_{A,B} = R_{A,B} e^{i\phi_{A,B}}$ into Eq.~(\ref{eq:S9}) and dividing it by $e^{i\phi_{A,B}}$ yields

\begin{eqnarray}
\label{eq:S16}
\dot{R}_{A,B} + iR_{A,B}\dot{\phi}_{A,B} + R_{A,B}\Big[ \frac{i\omega}{2} - \frac{ib_1}{2\omega} &+ \dfrac{b_0 + K_{\tau c}\sin^2(kx_c)}{2} - \dfrac{i\sigma}{2\omega}e^{-i\omega \tau_h}\Big] - \dfrac{3i\sigma_3}{8\omega}R^3_{A,B}e^{-i\omega\tau_h} \nonumber\\ &- \dfrac{K_{\tau c}\sin^2(kx_c)}{2}R_{B,A}e^{-i\omega\tau_c}e^{-i(\phi_{B,A}-\phi_{A,B})} = 0.
\end{eqnarray}
Expressing exponential terms as sines and cosines, and separating the real and imaginary parts gives us the temporal variation of the amplitude and phase of the oscillators as follows:
\begin{align} \label{eq:S17}
&\dot{R}_{A,B} + R_{A,B}\Big[ \frac{b_0 + K_{\tau c}\sin^2(kx_c)}{2} - \frac{\sigma}{2\omega}\sin(\omega \tau_h)\Big] - \frac{3\sigma_3}{8\omega}R^3_{A,B}\sin(\omega\tau_h) \nonumber\\ &\quad - \frac{K_{\tau c}\sin^2(kx_c)}{2}R_{B,A} [\cos(\omega\tau_c)\cos(\phi_{B,A}-\phi_{A,B})  + \sin(\omega\tau_c)\sin(\phi_{B,A}-\phi_{A,B})] = 0,\\
& R_{A,B}\dot{\phi}_{A,B} + R_{A,B}\Big[ \frac{\omega}{2} - \frac{b_1}{2\omega} - \frac{\sigma}{2\omega}\cos(\omega \tau_h)\Big] - \frac{3\sigma_3}{8\omega}R^3_{A,B}\cos(\omega\tau_h) \nonumber\\ &\quad + \frac{K_{\tau c}\sin^2(kx_c)}{2}R_{B,A} [\sin(\omega\tau_c)\cos(\phi_{B,A}-\phi_{A,B})  - \cos(\omega\tau_c)\sin(\phi_{B,A}-\phi_{A,B})] = 0.
\label{eq:S18}
\end{align}
On simplifying Eqs.~(\ref{eq:S17}) and (\ref{eq:S18}) using trigonometric identities, we get
\begin{align} \label{eq:S19}
\dot{R}_{A,B} + R_{A,B}\Big[ \frac{b_0 + K_{\tau c}\sin^2(kx_c)}{2} &- \frac{\sigma}{2\omega}\sin(\omega \tau_h)\Big] - \frac{3\sigma_3}{8\omega}R^3_{A,B}\sin(\omega\tau_h) \nonumber\\ &- \frac{K_{\tau c}\sin^2(kx_c)}{2}R_{B,A}\cos[\omega\tau_c - (\phi_{B,A}-\phi_{A,B})] = 0,\\
\dot{\phi}_{A,B} + \Big[ \frac{\omega}{2} - \frac{b_1}{2\omega} &- \frac{\sigma}{2\omega}\cos(\omega \tau_h)\Big] - \frac{3\sigma_3}{8\omega}R^2_{A,B}\cos(\omega\tau_h) \nonumber\\ &+ \frac{K_{\tau c}\sin^2(kx_c)R_{B,A}}{2R_{A,B}} \sin[\omega\tau_c - (\phi_{B,A}-\phi_{A,B})] = 0.
\label{eq:S20}
\end{align}

Let us denote the phase difference between the oscillators, $\phi_B -\phi_A$, as $\theta$. Using Eq.~(\ref{eq:S20}), we derive the equation for $\dot{\theta} = \dot{\phi}_B -\dot{\phi}_A$ as below:
\begin{align} \label{eq:S21}
\dot{\theta} - \frac{3\sigma_3}{8\omega}(R^2_B - R^2_A)\cos(\omega\tau_h) + \frac{K_{\tau c}\sin^2(kx_c)}{2}\Big[\frac{R_A}{R_B} \sin(\omega\tau_c + \theta) - \frac{R_B}{R_A} \sin(\omega\tau_c - \theta)\Big] = 0.
\end{align}
Assuming that the oscillators have similar amplitudes, i.e., $R_A = R_B = R$, the above equation simplifies to the following:
\begin{align} \label{eq:S22}
\dot{\theta} &= -K_{\tau c}\sin^2(kx_c)\cos(\omega \tau_c) \sin(\theta).
\end{align}
Consequently, simplifying Eqs.~(\ref{eq:S19}) and (\ref{eq:S20}) using the same assumption of similar amplitude, we get the following slow flow equations:
\begin{align} \label{eq:S23}
\dot{R} &= R\Big[ \frac{\sigma}{2\omega}\sin(\omega \tau_h) - \frac{b_0}{2} - \frac{K_{\tau c}\sin^2(kx_c)}{2}(1 + |\cos(\omega \tau_c)|) \Big] + \frac{3\sigma_3}{8\omega}R^3\sin(\omega\tau_h),\\
\label{eq:S24}
\dot{\phi}_{A,B} &= -\frac{\omega}{2} +\frac{b_1}{2\omega} + \frac{\sigma}{2\omega}\cos(\omega \tau_h) + \frac{3\sigma_3}{8\omega}R^2\cos(\omega\tau_h) - \frac{K_{\tau c}\sin^2(kx_c)}{2} \sin(\omega\tau_c).
\end{align}

Equation~\ref{eq:S23} predicts the occurrence of subcritical Hopf bifurcation since the coefficient of the cubic nonlinearity is positive. To approximately determine the frequency $f$ of the oscillations in the coupled system, we ignore the amplitude dependent term in Eqs.~(\ref{eq:S24}) and compute $\omega + \dot{\phi}_{A,B}$. To predict the occurrence of amplitude death in delay coupled non-identical oscillators, we consider the governing equations Eqs. (18) and (19) of the main manuscript and follow the steps as outlined by Eqs.~(\ref{eq:S1})-(\ref{eq:S14}) with the additional assumption of small values of mismatch parameter $\alpha$ (i.e., $1/(1+\alpha) \approx 1 - \alpha$, $\sin(kx\alpha) \approx kx\alpha$ and $\cos(kx\alpha) \approx 1$). This gives us the condition for amplitude death in delay coupled non-identical Rijke tube oscillators as:
\begin{align}
\left|\cos\left(\omega\tau_c\right)\right|< \Bigg[1 - \frac{(\sigma/\omega) \sin\left(\omega\tau_h\right)-b_0}{K_{\tau c}\sin^2(kx_c)} 
+\alpha\left(\frac{(\beta+\epsilon)\sigma/\omega \sin\left(\omega\tau_h\right)-\epsilon b_0}{2K_{\tau c}\sin^2(k x_c)} \right) \Bigg],
\label{eq:20}
\end{align}
where $\beta=1+2kx_f\cot(2kx_f)$ and $\epsilon = 3/2 - k x_c \cot(k x_c)$.

\bibliography{SupplementaryMat}